\UseRawInputEncoding
\documentclass[%
 reprint,
 amsmath,amssymb,
 aps,
 nofootinbib,
]{revtex4-2}

\usepackage{textcomp}
\usepackage{hyperref}
\usepackage{float}
\usepackage[english]{babel}  % Correct
\usepackage[utf8]{inputenc}  % si pdfLaTeX
\usepackage[T1]{fontenc}
\usepackage{multirow} 
\usepackage{tabularx}
\usepackage{multirow}
\usepackage{booktabs,multirow,siunitx}
\usepackage{graphicx}% Include figure files
\usepackage{dcolumn}% Align table columns on decimal point
\usepackage{bm}% bold math
\usepackage{orcidlink}
\usepackage{booktabs}
\begin{document}

\preprint{APS/123-QED}

\title{UV Luminosity Functions from HST and JWST: A Possible Resolution to the High-Redshift Galaxy Abundance Puzzle and Implications for Cosmic Strings}% Force line breaks with \\

\author{Mattéo Blamart\,\orcidlink{https://orcid.org/0009-0009-1001-9761}}
\email{matteo.blamart@mail.mcgill.ca}
 \affiliation{Department of Physics, McGill University, 3600 Rue University, Montreal, QC H3A 2T8, Canada \\
and Trottier Space Institute, 3550 Rue University, Montreal, QC H3A 2A7, Canada}%Lines 

\author{Adrian Liu\,\orcidlink{0000-0001-6876-0928}}
 \email{adrian.liu2@mcgill.ca}
\affiliation{Department of Physics, McGill University, 3600 Rue University, Montreal, QC H3A 2T8, Canada \\
and Trottier Space Institute, 3550 Rue University, Montreal, QC H3A 2A7, Canada}%

\author{Robert Brandenberger\,\orcidlink{0000-0001-7194-5691}}
\email{rhb@physics.mcgill.ca}
\affiliation{Department of Physics, McGill University, 3600 Rue University, Montreal, QC H3A 2T8, Canada \\
and Trottier Space Institute, 3550 Rue University, Montreal, QC H3A 2A7, Canada}%

\author{Julian B. Mu\~noz\,\orcidlink{https://orcid.org/0000-0002-8984-0465}}
\email{julianbmunoz@utexas.edu}
\affiliation{
 Department of Astronomy, The University of Texas at Austin, 2515 Speedway, Stop C1400, Austin, TX 78712, USA}
\affiliation{Cosmic Frontier Center, The University of Texas at Austin, Austin, TX 78712}
\affiliation{Texas Center for Cosmology \& Astroparticle Physics, Austin, TX 78712}

\author{Bryce Cyr\,\orcidlink{https://orcid.org/0000-0002-6569-3093}}
\email{brycecyr@mit.edu}
\affiliation{Center for Theoretical Physics, A Leinweber Institute,
Massachusetts Institute of Technology, Cambridge, MA 02139,USA}

\date{\today}% It is always \today, today,

\begin{abstract}

Recent observations of high redshift galaxies by the James Webb Space Telescope suggest the presence of a bright population of galaxies that is more abundant than predicted by most galaxy formation models. These observations have led to a rethinking of these models, and numerous astrophysical and cosmological solutions have been proposed, including cosmic strings, topological defects that may be remnants of a specific phase transition in the very early moments of the Universe. In this paper, we integrate cosmic strings, a source of nonlinear and non-Gaussian perturbations, into the semi-analytical code \texttt{Zeus21}, allowing us to efficiently predict the ultraviolet luminosity function (UVLF). We conduct a precise study of parameter degeneracies between star-formation astrophysics and cosmic-string phenomenology. Our results suggest that cosmic strings can boost the early-galaxy abundance enough to explain the measured UVLFs from the James Webb and Hubble Space Telescopes from redshift $z=4$ to $z=17$ without modifying the star-formation physics. In addition, we set a new upper bound on the string tension of $G\mu \lessapprox 10^{-8}$ ($95\%$ credibility), improving upon previous limits from the cosmic microwave background. Although with current data there is some level of model and prior dependence to this limit, it suggests that UVLFs are a promising avenue for future observational constraints on cosmic-string physics.

\end{abstract}

\maketitle

%\tableofcontents

\section{Introduction}
The ultraviolet luminosity function (UVLF) has become an important source of information for understanding the formation of the very first galaxies and the evolution of their properties from Cosmic Dawn to the present day \cite{Finkelstein_2016}. UVLFs measure the abundance or number density of these galaxies as a function of their UV luminosity or magnitude. This observable can be used to test astrophysics and cosmology, including possible deviations from the standard $\Lambda \text{CDM}$ cosmological model \citep{Trenti_2010,Ceverino2017,Tacchella_2018,Mirocha_2016,Park_2019,Mason_2019,Corasaniti_2017,Menci_2018,Sabti_2022,Rudakovskyi_2021,Sabti_2022,Sabti_2023,Urrutia_2025}.

The Hubble Space Telescope (HST) has measured these UVLFs up to redshift $z\approx 10$ \citep{Bouwens_2021,Livermore_2017,Atek_2015,Bouwens_2015}, and recently the James Webb Space Telescope (JWST) has pushed back this redshift barrier using photometric measurements up to redshift $z=17$ \citep{Finkelstein_2022,Castellano_2022,Bouwens_2023,kokorev2025,kokorev_2025} (with some samples also spectroscopically confirmed \citep{harikane2023,Arrabal_Haro_2023,Curtis-LakeSpectro,Wang_2023,Harikane_2025}) and potentially to the $z\approx17$ to $30$ redshift range \cite{perezgonzalez2025,castellano2025}. These new results suggest that UV bright galaxies are far more abundant at high redshifts than expected or predicted by most galaxy formation models and simulations, as well as the various extrapolations based on observations from lower redshifts \cite{Tacchella_2018,Yung_2018,behroozi_universe_2020,vogelsberger_high-redshift_2020,kannan_thesan_2022,kannan_millenniumtng_2023}. These results seem to highlight a radical change in UVLFs during the transition from HST to JWST observations, which needs to be understood. A wide range of explanations have been proposed in the literature, some based on modifying the astrophysics of galaxy formation and others on modifying $\Lambda \text{CDM}$ cosmology. On the astrophysical side, one explanation is a large enhanced star-formation/UV emission at high redshift while another is to assume a large stochastcity in the UV brightness of early galaxies \citep{inayoshi_lower_2022,finkelstein_ceers_2023,steinhardt_templates_2023,harvey_epochs_2024,yung_are_2024,dekel_efficient_2023,finkelstein_complete_2024,ceverino_redshift-dependent_2024,chworowsky_evidence_2024,shen_impact_2023,sun_seen_2023,sun_bursty_2023,pallottini_stochastic_2023,munoz_breaking_2023,casey_cosmos-web_2024,Wang_2024,Hutter_2025,Dhandha_2025,Chakraborty_2026,Gelli_2024,munoz2026relativelyfastreasonablyfurious}. Other explanations suggest the presence of an initial high stellar mass function, cosmic variance and/or sampling bias in the first JWST observations of a single field, as well as an absence of dust attenuation at ultra-high redshifts \citep{ferrara_stunning_2023,mason_brightest_2023,desprez_cdm_2024,willott_steep_2024}. Several solutions from a cosmological angle have also been proposed, such as the presence of primordial black holes, early or modified dark energy models, modified gravity, nonstandard dark matter, even a modification of the $\Lambda \text{CDM}$ model or even modifications to the $\Lambda \text{CDM}$ parameters themselves. \citep{menci_high-redshift_2022,biagetti_high-redshift_2023,gong_fuzzy_2023,hutsi_did_2023,shen_early_2024,padmanabhan_alleviating_2023,fakhry2025,Chakraborty_anti_2026,Jiang_2025,Adil_2023,Sokoliuk_2025,moffat2024galaxyformationearlyuniverse,Davari_2024}. Finding a cosmological solution is a challenge, given that the deviations from $\Lambda \text{CDM}$ model are not only severely constrained by observations such as the cosmic microwave background (CMB) and galaxy surveys at lower redshifts, but also by the UVLFs measured by HST. For example, in Ref. \cite{sabti_insights_2024}, it is shown that an enhancement of the matter power spectrum could explain the JWST observations, but would disagree with the HST data.

In this paper, we explore in detail models that involve cosmic strings as a possible solution \citep{jiao_early_2023,jiao_n-body_2024,cyr_not-quite-primordial_2025,shlaer_early_2012,jiao_accretion_2024,koehler_investigating_2024,olum_reionization_2006}. Generally speaking, cosmic strings are linear topological defects resulting from a specific phase transition in the very first instants of the Universe, and are predicted by many Grand Unification Theory models \citep{brandenberger_topological_1994,hindmarsh_cosmic_1995,vilenkin_cosmic_2000,durrer_cosmic_2002,kibble_implications_1980,kibble_phase_1982}. Compared to changes in the matter power spectrum, cosmic strings represent a qualitatively different way to modify $\Lambda \text{CDM}$ because of their ability to boost the formation of dark matter haloes (and therefore the production of bright galaxies) at high redshift, while at lower redshift, their contribution gradually becomes negligible. This suggests cosmic strings as a well-motivated explanation of UV bright high-redshift galaxies consistent with both JWST and HST observations, without sudden changes in astrophysical properties of galaxies. At ultra-high redshift, the $\Lambda \text{CDM}$ model predicts an exponential suppression of high-mass dark matter haloes. Cosmic strings have the ability to generate nonlinear dark matter halos at arbitrarily early times, resulting in
a only mild (power law as function of redshift) suppression of the nonlinear halo mass function.
The idea is that bright objects at ultra-high redshift originate from cosmic string-seeded dark matter haloes, while retaining the astrophysical properties of galaxies at lower redshift.

\begin{figure}
\includegraphics[width=\linewidth]{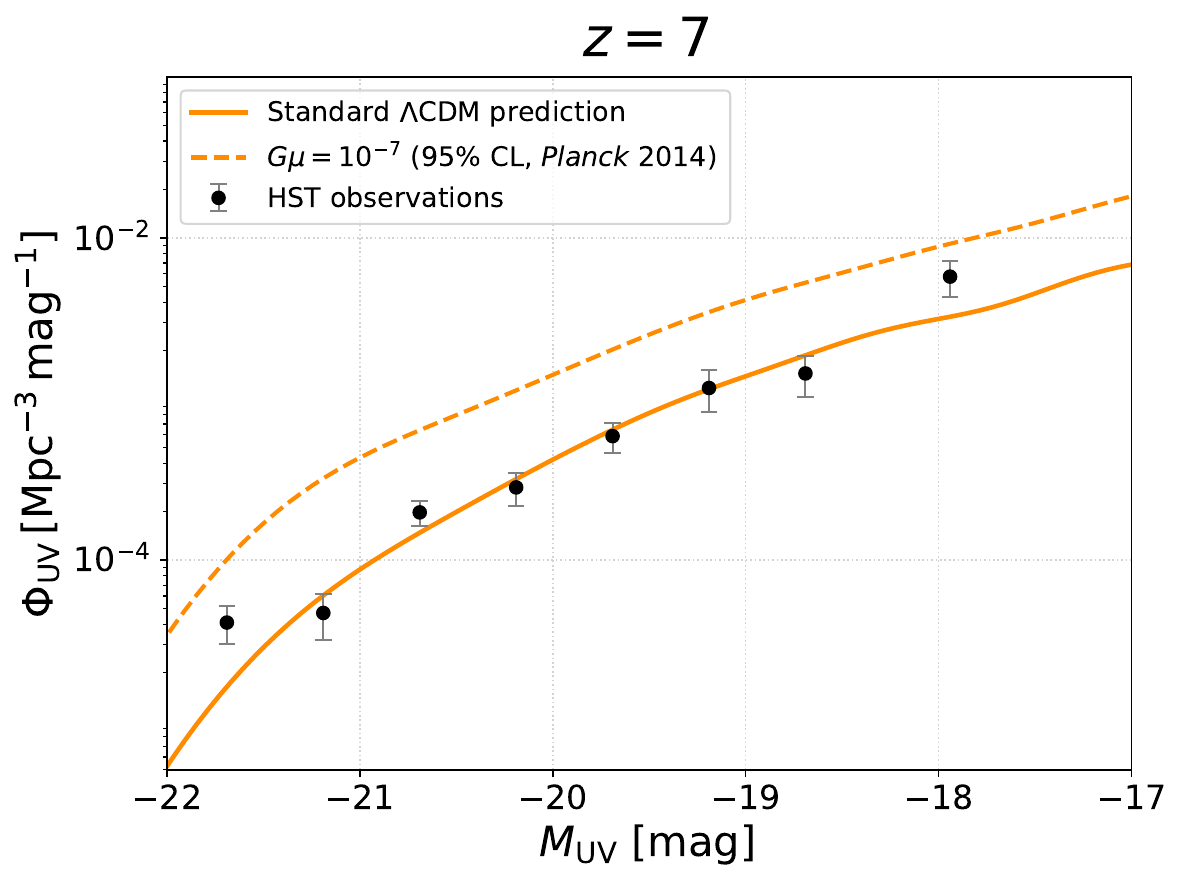}
\caption{\label{fig:illustration} Comparison between predictions of UVLFs with (dashed line) and without (solid line) the presence of cosmic strings with a string tension corresponding to the $95\%$ credibility limits (CL) from \textit{Planck} 2014. HST observations from Ref. \cite{Bouwens_2021} are shown in comparison in black with their error bar corresponding to a $68\%$ confidence interval. In both cases, the astrophysical parameters are standard for galaxy formation models and are the same in both predictions; only the presence of strings changes. These results highlight the very high sensitivity of UVLFs to the presence of cosmic strings for string tensions close to the limits obtained by \textit{Planck} data. This sensitivity is greater at higher redshifts.}
\end{figure}

Recently, numerous analytical studies, $N$-body simulations and also large-volume hydrodynamical simulations, have confirmed the impact of cosmic strings on the formation of bright objects and their potential as a viable solution to explain the abundance of early galaxies \citep{jiao_early_2023,jiao_n-body_2024,koehler_investigating_2024}. These cosmic strings are in the form of a vast network with a dominant portion of loops and some long lines. The stable loops are characterized by an energy per unit of length $\mu$ (where $G\mu$ refers to the dimensionless string tension in natural units
with the gravitational constant $G=6.7 \times 10^{-39} \rm{GeV}^{-2}$). They have the capacity to accrete matter and thus trigger the formation of nonlinear haloes at arbitrarily high redshift. The more energetic the strings, the stronger their impact.

However, it is extremely difficult to study the interplay between different possible astrophysical scenarios and cosmological scenarios with cosmic strings. The difficulties are that theoretical analyses and $N$-body simulations tend to predict a distribution of dark matter haloes from which galaxies will form. UVLFs, on the other hand, give no direct access to information on the mass of these haloes, how the accreted baryons are converted into stars, or the quantity of light per unit mass emitted by these stars in the UV. The optimal solution would be to test all these scenarios with large-volume hydrodynamical simulations, but their long computation times make this impossible for the time being. 
Instead, we use the semi-analytic code \texttt{Zeus21}\footnote{\url{https://github.com/JulianBMunoz/Zeus21}} to overcome these problems, providing very rapid (millisecond) predictions for UVLFs given a model \citep{cruz_effective_2025,munoz_breaking_2023,munoz_effective_2023}. The UVLF module is also flexible enough to test different scenarios and interplays, both cosmological (with a CLASS-based infrastructure \cite{lesgourgues_cosmic_2011}) and astrophysical (with a highly flexible halo-galaxy connection). We have incorporated cosmic string predictions for the halo mass function in \texttt{Zeus21} to rapidly predict UVLFs with numerous astrophysical scenarios.

In this work, we address two specific objectives. The first is to test more precisely the extent to which cosmic strings are a viable solution to reconcile the JWST and HST data. 
Secondly, we present UVLFs in detail as a new observing window for studying cosmic string loops and obtaining new constraints. The most robust constraint on the string tension comes from CMB anisotropies and stipulates that $G\mu\le10^{-7}$ at $95\%$ the credibility level \citep{charnock_cmb_2016,Planck_2014,dvorkin_cosmic_2011} \footnote{A recent study presents updated constraints on cosmic strings using data from the Atacama Cosmology Telescope Data Release 6 and sets an upper limit of $G\mu\le4 \times10^{-8}$ at $95\%$ credibility level \citep{raidal2026}}. The sensitivity of UVLFs to cosmic strings is shown in Figure \ref{fig:illustration}, with a comparison between a standard UVLF prediction and one that includes the extra contributions of cosmic strings with a string tension equal to the upper bound obtained from \textit{Planck} 2014 data. Observations of UVLFs by HST show that UVLFs are likely to improve constraints compared to just the CMB.
Recent evidence of a gravitational wave background from several millisecond pulsar timing arrays has led to a potential limit of $G\mu\le10^{-10}$ \citep{blanco-pillado_comparison_2021,afzal_nanograv_2023,Abbott_2021,Hindmarsh_2023,Kume_2024}. This constraint may be less robust, however, as it is highly dependent on the uncertain modeling of the size distribution of the small loops that dominate gravitational wave production, as well as on the way these waves are emitted \cite{Abbott_2021,Hindmarsh_2023,Kume_2024}. This modeling uncertainty may also lead to a less stringent upper bound on the tension of cosmic strings, bringing it closer to the value obtained from analyses of CMB data \cite{Hindmarsh_2023,Kume_2024}. For UVLFs, small loops have very little impact and it is the large loops that act as additional gravitational seeds in the early universe alongside the density fluctuations described by standard $\Lambda \text{CDM}$ cosmology. Since the abundance of large loops is set by causality arguments, predictions relying only on large loops are more robust.
Obtaining a new limit on $G\mu$ is important because it provides information about the physics of the primordial universe and possible phase transitions. The reason is that $\mu$ is related to the energy scale $\eta$ of the string-forming phase transition with $\mu\approx \eta^2$ in natural units. An improved upper bound can rule out more models of high-energy physics in particular Grand Unification Theories.

The rest of this paper is organized as follows. Section~\ref{sec:methods} presents the paper's methodology both in terms of our modelling of cosmic strings and our modelling of high-redshift galaxies. We provide some intuition for how cosmic strings affect the high-redshift universe and how there can be degeneracies between cosmic string phenomonology and the astrophysics of galaxies. Section~\ref{sec:results} presents our results including an array of new limits on $G\mu$. Section~\ref{sec:discussion} contains our discussion of these results, highlighting the assumptions, limitations, and future prospects of using UVLFs to study constrain cosmic strings. We summarize our conclusions in Section~\ref{sec:conclusions}.
\section{Methods}
\label{sec:methods}
%Appuyer sur le fait que normalement sur les plus grandes loops les prédictions sont assez similaires
In this section, we will detail all the necessary steps, starting with a modification of the halo mass function related to cosmic strings, its incorporation into \texttt{Zeus21}, and ending with specific predictions of UVLFs linking to the data.

\subsection{Halo mass function seeded by string loops}
\label{sec:hmffromstrings}
Several models can be used to describe the loop size distribution of cosmic string loops \citep{austin_evolution_1993,albrecht_evolution_1985,bennett_evidence_1988,allen_cosmic-string_1990,vanchurin_scaling_2006,lorenz_cosmic_2010,ringeval_cosmological_2007,blanco-pillado_large_2011,blanco-pillado_number_2014,auclair_cosmic_2019,blanco-pillado_direct_2020}. Recent theoretical and numerical works have refined and confirmed the effects between structure formation and string loops assuming a one-scale model, which we choose as our reference model \citep{copeland_scaling_1992,perivolaropoulos_cobe_1993,austin_evolution_1993}. Nevertheless, this work is theoretically generalizable and applicable to any other model describing the distribution of cosmic string loops.\\

The one-scale model assumes that cosmic string loops formed at time $t$ have radius $R=\alpha t$, with constant $\alpha\approx0.1$. The loops are not perfectly circular and their length $l$ is assumed to be $l=\beta R$ with $\beta \approx 10$. The distribution is also dependent on and influenced by the average number of $N$ infinite strings passing through the Hubble volume. In theory, the parameters $N,\alpha$ and $\beta$ are determined by the physics governing the evolution of the string network. Although analytical work shows that the string loop distribution over time follows a scaling solution, it does not currently predict a specific value of these three coefficients. The value of these three coefficients is therefore obtained using numerical simulation \citep{ringeval_cosmological_2007,vanchurin_scaling_2006,lorenz_cosmic_2010,blanco-pillado_large_2011,blanco-pillado_number_2014,blanco-pillado_direct_2020,auclair_cosmic_2019}. These simulations, assume an effective description of the strings by the Nambu-Goto action (which neglects the effects of the string core) and the coefficients obtained are of the order of $\alpha=0.1$, $\beta=10$ and $N=570$. For the rest of this work, we have chosen to keep these values as constants for $\alpha$, $\beta$ and $N$ but this analysis can also be generalized considering $\alpha$ and $\beta$ as free parameters of the model and here the final results are entirely dependent on the pair $G\mu$ and $N$. In the one-scale model, the distribution of loops and their density is described as a function of their size. The number density $n(R,t)dR$ (in comoving coordinates) of
loops in the radius interval between $R$ and $dR$ at time $t$ is
\begin{equation} \label{eq:loopdistribution}
n(R, t) = 
\begin{cases}
N \alpha^2 \beta^{-2} t_0^{-2} R^{-2} &  \alpha t_{\text{eq}} < R \leq \alpha t \\
N \alpha^{5/2} \beta^{-5/2} t_{eq}^{1/2} t_{\text{0}}^{-2} R^{-5/2} & R_c(t_\text{eq}) \leq R \leq \alpha t_{\text{eq}} \\
n(R_c(t), t) &  R < R_c(t_\text{eq})
\end{cases}
\end{equation}
with $R_c(t) \equiv \gamma \beta^{-1} G\mu t$, $t_0$ the present time, $t_{eq}$ the time of matter-radiation equality and $\gamma$ a constant determined by the strength of gravitational wave emission per string vibration which is of the order of $10^2$ \cite{vachaspati_gravitational_1985}. This gravitational wave emission is accompanied by a reduction in the radius of the loops. In this equation, the first line corresponds to the loops formed during the matter era. The second line corresponds to loops formed during the radiation domination era but with a size large enough for the loop radius decay to be negligible, whereas for the third line, these smaller loops have a very short lifetime of less than one Hubble time and have a negligible effect on structure formation. For this reason, the exact distribution of these small loops is of little importance and we will assume $n(R,t)$ to be constant for the smaller loops in this range of radii in this analysis.\\ 

Now that we have presented the distribution of cosmic string loops, we consider how these loops lead to the creation of dark matter haloes. The following derivation follows the steps described in Refs \citep{jiao_early_2023,jiao_n-body_2024}.  In fact, since loops are overdensities, they will accrete matter. Moreover, the average separation of loops is larger than the region through which a cosmic string loop accretes matter. Hence, it is a good approximation to consider the accretion of matter by one loop as independent of the other loops. 
In this case, at a given time $t$, a cosmic string loop of radius $R$ will have accreted a mass $M(R,t)$. The halo mass function $dn/dM_h |_{\rm{CS}}$ seeded by cosmic string loops is then derived from $n(R,t)$ above, the relation (\ref{eq:loopdistribution}) and the following relation
\begin{equation} \label{halomassloop}
    \dfrac{dn}{dM_h}\Bigg|_{\rm{CS}}=n(R(M),t)\dfrac{dR}{dM}.
\end{equation} In order to calculate the final loop-seeded nonlinear halo mass, it is necessary to relate the mass accreted by a loop to its loop radius where $R(M)$ is the inverse function of $M(R,t)$, which is given by
\begin{equation}
M(R, z) = 
\begin{cases}
\beta \mu R \left( \dfrac{z_{\text{f}} + 1}{z + 1} \right) &\alpha t_{\text{eq}} < R \leq \alpha t \\
\beta \mu R \left( \dfrac{z_{\text{eq}} + 1}{z + 1} \right) &R_c(t_{eq}) \leq R \leq \alpha t_{\text{eq}} \\
\beta \mu \dfrac{R^2}{R_c(t_{\text{eq}})} \left( \dfrac{z_{\text{eq}} + 1}{z + 1} \right) & R < R_c(t_{eq})
\end{cases}
\end{equation}
with $z_\text{f}$ the redshift when the loop formed and $z_\text{eq}$ the redshift of the equal matter-radiation time where we are switching between redshift and time for convenience, assuming standard $\Lambda \text{CDM}$ relations with cosmological parameters taken from. This equation illustrates the fact the mass varies as $(1+z)^{-1}$. For the third line, given the short lifetime of these loops, they have a highly suppressed mass accretion symbolised by the ratio $R^2/R_c(t_\text{eq})$.
For this model, the final halo mass function seeded by string loops is as follows. First, we define several characteristic masses for notational convenience:

\begin{align}
M_u(z)            &\equiv 4.5 \times 10^7 M_\odot 
    \left( \frac{G\mu}{10^{-10}} \right)
    \left( \frac{z_\text{eq}+1}{z+1} \right)^{3/2}, \label{eq:Mu} \\[4pt]
M_{\text{eq}}(z)  &\equiv 1.98 \times 10^8 M_\odot 
    \left( \frac{G\mu}{10^{-10}} \right)
    \left( \frac{z_\text{eq}+1}{z+1} \right), \label{eq:Meq} \\[4pt]
M_{\text{min}}(z) &\equiv 0.45 M_\odot 
    \left( \frac{G\mu}{10^{-10}} \right)^2
    \left( \frac{z_\text{eq}+1}{z+1} \right) \label{eq:Mmin}
\end{align}
with $M_\odot $ being the solar mass. 
\vspace{0.5em}

\noindent Now the final halo mass function seeded by string loops is

\begin{widetext}
\begin{equation} 
\dfrac{dn}{dM_h}\Bigg|_{\rm{CS}} =
\begin{cases}
0, & M > M_u(z), \\[4pt]
2.6 \times 10^{25} \dfrac{N}{(1+z)^3 M^4}
\left( \dfrac{G\mu}{10^{-10}} \right)^3, & M_{\text{eq}}(z) < M \leq M_u(z), \\[4pt]
4.7 \times 10^7 \dfrac{N}{(1+z)^{\frac{3}{2}} M^{\frac{5}{2}}}  
\left( \dfrac{G\mu}{10^{-10}} \right)^{3/2} , & M_{\min}(z) < M \leq M_{\text{eq}}(z), \\[4pt]
4.7 \times 10^7 \dfrac{N}{(1+z)^{\frac{3}{2}}M_{\min}^{\frac{5}{2}}}
\left( \dfrac{G\mu}{10^{-10}} \right)^{3/2}, & M \leq M_{\min}(z).
\end{cases}
\label{eq:halomassloop}
\end{equation}
\end{widetext}
with $M$ in solar masses.
These four regimes of the halo mass function are related to the three size regimes of cosmic string loops in the one scale model described above. Here the first regime indicates that there are no loops larger than $M_u(z)$ and therefore no dark matter haloes seeded by these loops. This corresponds to a cut-off in the loop size of the one-scale model. \\

In connection with the $\Lambda \text{CDM}$ model and the \texttt{Zeus21} code, the assumed standard halo mass function is that of Sheth $\&$ Tormen (henceforth ST) \cite{sheth_excursion_2002}. As demonstrated analytically and confirmed by simulation by Ref. \citep{jiao_early_2023,jiao_n-body_2024}, in a universe containing cosmic string loops, the halo mass function can be written as the sum of halos formed by fluctuations in the $\Lambda \text{CDM}$ model and seeded halos by cosmic string loops: 
\begin{equation} \label{eq:halomasstot}
\dfrac{dn}{dM_h} \Bigg|_{\rm{Total}}\approx\dfrac{dn}{dM_h} \Bigg|_{\rm{ST}}+\dfrac{dn}{dM_h} \Bigg|_{\rm{CS}}
\end{equation}
with $dn/dM_h \big|_{\rm{ST}}$ corresponding to the ST halo mass function and $dn/dM_h \big|_{\rm{CS}}$ the halo mass function seeded by string loops Equation \ref{eq:halomassloop}.
\begin{figure*}
\includegraphics[width=\linewidth]{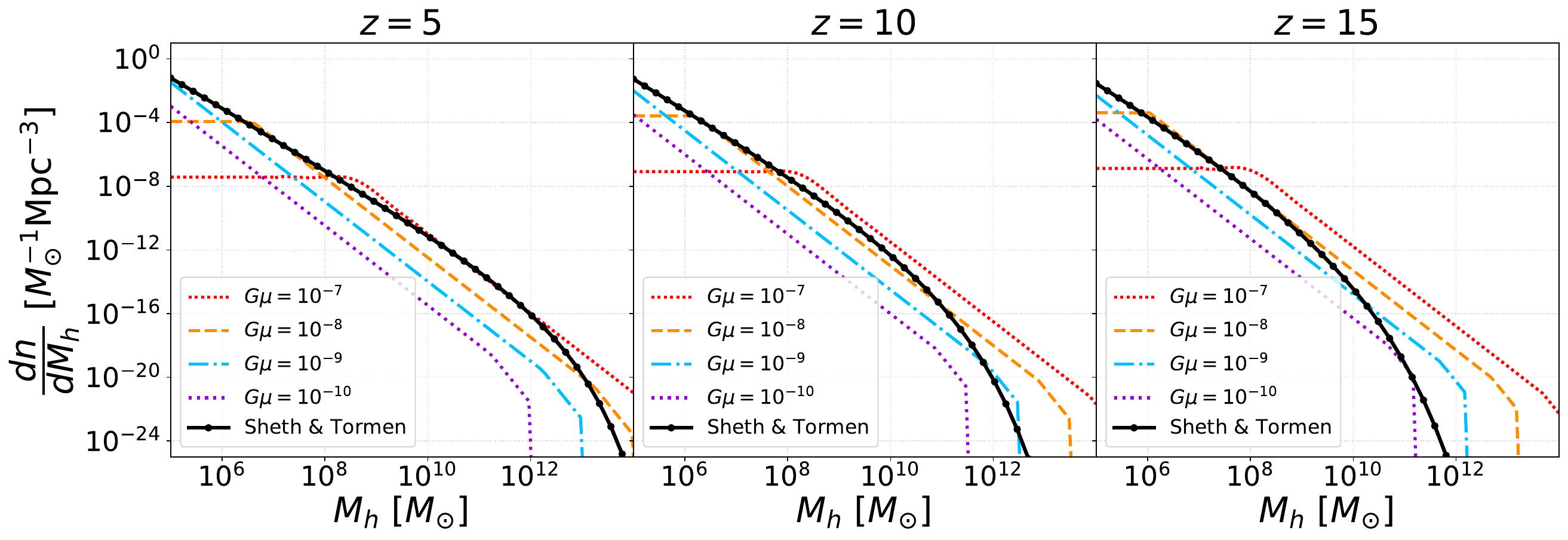}% Here is how to import EPS art
\caption{\label{fig:halomass} Comparison of the cosmic string loops halo mass function for different string tensions and the Sheth $\&$ Tormen (ST) halo mass function as a function of redshift. The factor $N$ in Equation \ref{eq:halomassloop} is set equal to $N=570$.  The vertical axis corresponds to the halo mass function and the horizontal axis to the mass range. As shown in Section~\ref{sec:hmffromstrings}, the cosmic string halo mass function becomes subdominant for most string tensions compared to the ST halo mass function. As redshift increases, the string halo mass function gradually tends to dominate the dark matter haloes high mass regime over the ST halo mass function. This figure highlights that cosmic strings can explain certain anomalies by forming dark matter haloes at high redshift while remaining consistent with observations at lower redshift when their contributions become subdominant.}
\end{figure*}
%Decription d'une figure avec la halo mass function loops
Figure \ref{fig:halomass} presents a comparison of the ST halo mass function and cosmic string loops as a function of redshift and different string tensions. As we can see, the greater the tension of the cosmic strings, the greater the rate of formation of dark matter halos. This is to be expected, since the more massive these strings are, the more they can boost matter accretion and the formation of dark matter haloes. The halo mass function of loops is increasingly dominant as a function of redshift, but also as a function of mass. In fact, the ST halo mass function predicts an exponential suppression of massive halos at higher redshifts and also shows an exponential decay as the mass increases, whereas the halo mass function of loops evolves as a $1/(1+z)^{3/2}$ as a function of redshift, and only has a power-law decay as the mass increases. At low redshift, the ST halo mass function completely dominates those of the loops so that  $dn/dM_h|_{\rm{Total}}\approx dn/ dM_h|_{\rm{ST}}$ for string tensions, in line with the upper bound from the CMB \citep{jiao_early_2023,jiao_n-body_2024}.

\subsection{Link between halo mass function and UV lumonisty function}
To study the impact of the modification of the halo mass function on the UVLF, we need a model that links the halos and UV brightness of the galaxies (as quantified by their UV magnitude $M_{UV}$) living in these haloes. With this in mind, we modify the \texttt{Zeus21} code, and more specifically the UVLF module. To model UVLFs, \texttt{Zeus21} assumes that all galaxies live in dark matter haloes with a halo occupation fraction of unity. This allows UVLFs to be modelled as 
\begin{equation}
    \phi_{UV}\equiv\dfrac{dn}{dM_{UV}}=\int dM_h \dfrac{dn}{dM_{h}} P(M_{UV}|M_h), 
\end{equation}
with $dn/dM_{h}$ the halo mass function and $P(M_{UV}|M_h)$ the probability that a halo of mass $M_h$ hosts a galaxy of magnitude $M_{UV}$. 
In the standard version of \texttt{Zeus21}, the halo mass function $dn/dM_{h}$ corresponds to ST halo mass function. Here, we implemented the halo mass function defined in Equation \ref{eq:halomasstot}.

\begin{figure*}[]
\includegraphics[width=\linewidth]{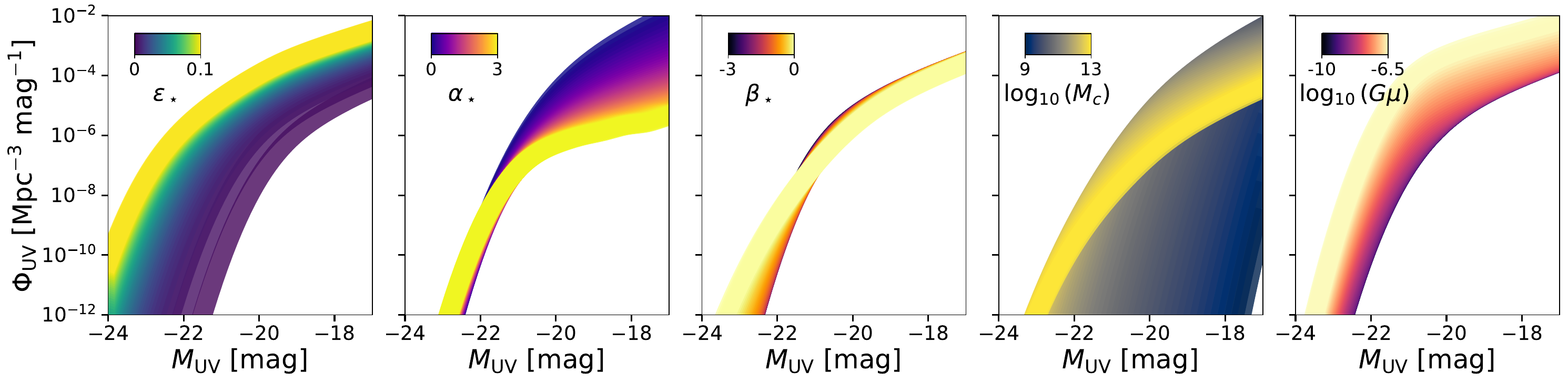}% Here is how to import EPS art
\caption{\label{fig:UVLF} Comparison of the impact of the five star formation parameters on final UVLF at redshift $z=8$. The first four parameters of the SFE and the last $G\mu$ related to cosmic strings (and $M_c$ in  units of $M_{\odot}$). Each parameter is varied individually while keeping all others constant. The factor $N$ in Equation \ref{eq:halomassloop} is set equal to $N=570$. The impact of cosmic strings tends to diminish and become completely negligible as $G\mu$ decreases.}
\end{figure*}

Although the ST halo mass function is a generic and clear prediction from the $\Lambda \text{CDM}$ model, the astrophysics of star formation is much more uncertain. However, by parametrizing astrophysical effects in a flexible way, it is possible to capture a range of possible behaviours and to understand the degeneracies between astrophysical and cosmological effects. The halo-galaxy connection is made through a function $P(M_{UV}|M_h)$ \cite{wechsler_connection_2018}. In \texttt{Zeus21}, it is a Gaussian centred around a predicted mean value $\overline{ M_{UV}}(M_h)$ and standard deviation $\sigma_{UV}$. In this study, $\sigma_{UV}$ will be set at its default value equal to $0.5$ \cite{munoz_breaking_2023}. The UV magnitude of a galaxy depends on the star formation rate $\dot{M}_\star$ (SFR) via 
 \begin{equation}\label{SFE}
 \dot M_\star=f_\star(z,M_h)f_b \dot M_h,
\end{equation} with $f_b\approx0.16$ the baryon fraction and \texttt{Zeus21} assumes an exponential accretion model $M_h(z)\propto e^{-a_{cc} z}$ with $a_{cc}=0.79$ \cite{aghanim_planck_2020,Schneider_2021}. The
average halo-galaxy connection is directly related to the star formation
efficiency (SFE) $f_{\star}(z,M_h)$ as a function of redshift $z$ and halo mass $M_h$. Several analytical studies and numerical simulations have shown that a double power-law function for the SFE, i.e., 
\begin{equation}
    f_{\star}=\dfrac{2\epsilon_{\star}}{(M_h/M_c)^{-\alpha_{\star}}+(M_h/M_c)^{-\beta_{\star}}}
\end{equation}
allows both numerical simulations and experimental data to be fitted correctly and to be flexible to different star formation scenarios \citep{moster_constraints_2010,furlanetto_minimalist_2017,sabti_galaxy_2022,madau_cosmic_2014}. This star formation efficiency function contains four free parameters. The first parameter $\epsilon_{\star}$ corresponds to the overall amplitude of the function, the parameters $\alpha_{\star}>0$ and $\beta_{\star}<0$ are the two powers of each power-law, $M_c$ corresponds to a critical mass at which we transition from the first power law with slope $\alpha_{\star}$ to the other with slope $\beta_{\star}$. The double power law reflects suppression in the SFE at low masses due to stellar, supernova and indirect reionization feedbacks, and at high masses supppresion due to feedback from active galactic nuclei. The SFR is then converted into a UV luminosity by the proportionality relation $L_{UV} = \dot M_{\star}/\kappa_{UV}$ with $\kappa_{UV}$ the conversion factor. We can therefore see that this factor $\kappa_{UV}$ is entirely degenerate with $\epsilon_{\star}$ and \texttt{Zeus21} therefore define an effective parameter $\epsilon_{\star,UV}\equiv\epsilon_{\star}(\kappa_{UV}/\overline{\kappa_{UV}})^{-1}$ for a fiducial $\overline{\kappa_{UV}} = 1. 15 \times 10^{-28} (M_{\odot}\ \text{yr}^{-1} )/(\text{erg s}^{-1} )$ as in Ref. \cite{madau_cosmic_2014}. The final conversion $\log_{10}(L_{UV}/(\text{erg s}^{-1}\text{Hz}^{-1}))=
0.4(51.63-M_{UV})$ converts $L_{UV}$ to $M_{UV}$ in the AB magnitude system.

The \texttt{Zeus21} UVLF module also takes into account dust attenuation, at the level of the data, which tends to reduce the apparent brightness of galaxies. Similar to the treatment presented in Ref. \cite{sabti_galaxy_2022}, \texttt{Zeus21} adopts the dust prescription calibrated by Ref. \cite{Meurer_1999} with the value measured in Ref. \cite{Bouwens_2015}. The \texttt{Zeus21} UVLF module extrapolates the dust attenuation to $z > 8$ by
using the $z = 8$ result, as advocated in Ref.  \cite{mason_brightest_2023}. In short, we have a flexible model for the SFR that is able to capture a wide spectrum of astrophysical scenarios through free parameters in the SFE, allowing us to study possible degeneracies between a cosmological and astrophysical scenario in the context of an explanation of UVLFs. 

Figure \ref{fig:UVLF} shows the impact of SFE parameters and string tension on the final UVLF at redshift $z=8$. Similar behaviour is obtained for other redshifts. The parameter $\epsilon_{\star}$, the amplitude of the SFE, leads to a large influence on the amplitude of the UVLF without necessarily changing the slope, whereas $\alpha_{\star}$ and $\beta_{\star}$ will respectively modify the UVLF slope in the low-luminosity (high $M_{UV}$) and high-luminosity regimes. The $M_c$ parameter will dictate the transition from the slope described by the $\alpha_{\star}$ parameter to that described by $\beta_{\star}$. Overall, the presence of string tension tends to shift the UVLF upwards. Furthermore, cosmic strings tend to boost the formation of haloes in the mass range where dark matter haloes are most likely to host a UV-bright galaxy.

\subsection{Scenarios tested and choice of priors}

In this work, the first objective is to find out whether it is possible to obtain a new upper bound on the string tension thanks to changes in the UVLFs when cosmic strings are introduced. It is also a study of the degeneracies between the possible impact of cosmic string loops and the various SFE scenarios. 
%For instance, numerous studies \citep{inayoshi_lower_2022,finkelstein_ceers_2023,steinhardt_templates_2023,harvey_epochs_2024,yung_are_2024,dekel_efficient_2023,finkelstein_complete_2024,ceverino_redshift-dependent_2024,chworowsky_evidence_2024,shen_impact_2023,sun_seen_2023,sun_bursty_2023,pallottini_stochastic_2023,munoz_breaking_2023,casey_cosmos-web_2024} propose a physical origin to the fact that the UVLFs measured by JWST at redshift $z>9$ could imply a much more efficient SFE than at redshift $z\leq8$. 
Secondly, the question is whether cosmic strings can explain these data both at high redshift measured by JWST and at lower redshift measured by HST, without necessarily having an abrupt change and increase in SFE.

\subsubsection{Conservative Scenario}

The first, very conservative scenario, similar to Ref. \cite{munoz_breaking_2023,sabti_galaxy_2022}, will look at each UVLF at each redshift independently of the others, with all UVLFs from both HST and JWST, assuming priors presented in Table \ref{tab:priors_scenarios} around each astrophysical parameter of the SFE and string tension $G\mu$. The astrophysical parameters are constrained independently at the different redshifts. The choices and ranges of priors are similar to those of other studies on the subject \citep{munoz_breaking_2023,sabti_galaxy_2022,yung_cdm_2025}. In summary, in this scenario, the astrophysical parameters are parameterized as follows:

\begin{align}
\label{eq:conservativeparam}
\textbf{Conservative} 
&= \Bigl\{ \text{Parameters } 
   \{\alpha_{\star}, \beta_{\star}, \epsilon_{\star}, M_c, G\mu \} \nonumber\\
&\quad \text{independent at each } z \Bigr\}
\end{align} In reality, $G\mu$ is a fixed constant. However, it is instructive to first consider $G\mu$ as an independent parameter for each redshift as an intuition-building exercise. One may then choose to subsequently combine the inferences at different redshifts to obtain one $G\mu$ constraint (while still keeping separate astrophysical parameters per redshift). This scenario is conservative in the sense that it discourages the presence of cosmic strings, as it allows the astrophysical parameters maximal flexibility to fit the observations. We observed that the upper bound on $G\mu$ for the conservative scenario depends slightly on the choice of the lower bound of the prior on $G\mu$. We therefore also included an additional choice of prior with a lower bound corresponding to the energy scale of the Large Hadron Collider (LHC), the minimum tension that cosmic strings can have, since cosmic strings have not been observed in these experiments. 

\subsubsection{Fiducial Scenario}
\label{sec:fidscenario}

In Refs. \cite{sabti_galaxy_2022,munoz_breaking_2023} it was shown (also verified as validation in this work) that the astrophysical parameters $\{\alpha_{\star},\space \beta_{\star}, \epsilon_{\star}, M_c\}$ evolve smoothly as a function of redshift. Physically, the SFE is nearly universal as a function of redshift for redshifts smaller than $z\le 9$. The parameter $\alpha_{\star}$ remains fairly constant as a function of redshift, while $\beta_{\star}$ cannot be precisely measured by HST's UVLF. The other two parameters $\epsilon_\star$ and $M_c$ evolve along as a power law as a function of redshift. It has been shown that assuming $\alpha_{\star},\beta_{\star}$ constant as a function of redshift and letting $\epsilon_\star$ and $M_c$ evolve in powerlaw as a function of redshift produces the best fit of the data with the fewest parameters, and this minimal or fiducial model is slightly preferred over the conservative model. This approach takes into account this quasi-universality property of the SFE as a function of redshift to reduce the number of free parameters.  The $G\mu$ parameter is always assumed to be constant, with log-uniform and uniform priors. Our fiducial model is therefore defined by the expressions
\begin{equation}
\label{eq:fiducialparam}
\textbf{Fiducial} =
\left\{
\begin{array}{l}
\alpha_{\star}(z) = \alpha_{\star} \\
\beta_{\star}(z) = \beta_{\star} \\
\log_{10} \epsilon_{\star}(z) = \epsilon_{\star}^s \times \log_{10} \left( \tfrac{1 + z}{1 + 6} \right) + \epsilon_{\star}^i \\
\log_{10} \left( \tfrac{M_c(z)}{M_\odot} \right) = M_c^s \times \log_{10} \left( \tfrac{1 + z}{1 + 6} \right) + M_c^i
\end{array}
\right.
\end{equation}
and thus our free parameters are $G\mu$, $\alpha_\star$, $\beta_\star$, $\epsilon_*^s$, $\epsilon_\star^i$, $M_c^s$, and $M_c^i$. Once again, the exact priors are listed in Table~\ref{tab:priors_scenarios}.

\subsection{Choice of Observation Samples}
\label{sec:samples}
First, for lower redshifts, we use UVLF constraints from Hubble Space Telescope (HST) obtained by Ref. \cite{Bouwens_2021} covering redshifts $z=4$ to $z=8$. Then at higher redshifts, we use UVLF constraints from the James Webb Telescope (JWST) obtained by various research groups. The UVLF used at redshifts $z=9$, $10$, and $11$ correspond to the results of Ref. \citep{finkelstein_ceers_2023} and Ref. \cite{donnan2024} while the UVLF used at redshifts $z=12$, $14$, and $17$ are a compilation of studies by different observations \citep{donnan2024,adams_epochs_2024-1,perez-gonzalez_life_2023,whitler_z_2025,robertson_earliest_2024,casey_cosmos-web_2024,castellano_pushing_2025,perez-gonzalez_rise_2025,franco_physical_2025,weibel_exploring_2025}. For these three redshifts, these are the same observation samples as those used in Ref.~\cite{yung_cdm_2025}. The HST observations are robust, making them an optimal choice for obtaining a new upper bound on the string tension. As for the JWST observations, although for the moment they are predominantly photometric, recent observations will continue to refine and confirm these results spectroscopically. Nevertheless, even photometrically, these results can also be used and interpreted as an upper bound on the UVLF \cite{castellano_pushing_2025}, making them once again similarly interesting data for obtaining a new upper bound on $G\mu$.

\subsubsection{Obtaining upper bounds}
To obtain upper bounds on $G\mu$, the conservative and fiducial scenarios presented in the subsection will be tested on different observational samples. For the fiducial scenario, we will use only HST data from $z=4$ to $z=8$. For the conservative scenario, the posterior distribution will be calculated individually on all observed HST and JWST redshifts from $z=4$ to $z=17$. We also included a computation of the marginal posterior distribution on $G\mu$ by combining the posterior distributions at each redshift from the conservative scenario. After marginalizing over the four astrophysical parameters independently, the marginal distributions at each redshift are then multiplied to compute the final marginal distribution. We combine the marginal posterior distributions of the conservative scenario for three different cases: HST data from $z=4$ to $z=8$, JWST data from $z=9$ to $z=17$, and combined HST and JWST data from $z=4$ to $z=17$.

In principle, the measured UVLFs are only sensitive to cosmic string loops formed 
during the radiation domination era (corresponding to the second line of Equation~\ref{eq:loopdistribution} and the third line of Equation~\ref{eq:halomassloop}) and therefore the results obtained on $G\mu$ are in fact constraints on the product $N (G\mu)^{3/2}$. Having set $N=570$ for this study and for simplicity of notation (and to conform with the custom in the cosmic string literature), we use the notation $G\mu$, but the results reported on the upper bounds and posterior distributions should be interpreted as constraints on $(N/570)^{2/3} G\mu$ rather than $G\mu$ alone.  

\subsubsection{Reconciling HST and JWST data}

For our second objective, which is to explain the JWST data without imposing a radical change in the SFE between the HST and JWST data, we will calculate the posterior distribution over all observed UVLFs from HST and JWST from $z=4$ to $z=17$ but considering the fiducial scenario. Indeed, the fiducial scenario imposes the quasi-universal nature of the SFE and its smooth evolution as a function of redshift.  
\subsection{Fitting and computing the posterior distribution on the UVLFS}

To calculate the posterior distribution, we assume a log likelihood at each redshift of the form
\begin{equation}
    -\log\mathcal{L}(z)=\sum_i \dfrac{(\phi_{UV,i}^{\text{obs}}-\phi_{UV,i}^{\text{bin}})^2}{2(\sigma_i^{\text{obs}})^2}
\end{equation}
with $\phi_{UV,i}^{\text{obs}}$ the UVLF measurement at a given redshift and at a given magnitude $M_{UV}$, $\phi_{UV,i}^{\text{bin}}$ the UVLF predicted using \texttt{Zeus21} code, $\sigma_i^{\text{obs}}$ the associated uncorrelated Gaussian errors and summing over the different magnitude bins $i$. We then sum the log likelihoods at different redshifts. To obtain the final constraints on both astrophysical and cosmological parameters, we use a Markov Chain Monte Carlo (MCMC) method with the \textit{emcee} package \cite{Foreman_Mackey_2013}. We have calculated the different posterior distributions using 120 walkers and 20,000 steps, discarding the first 500 as burn in \cite{yung_cdm_2025}. For redshift $z=17$ only, some measurements are only upper bounds, so for simplicity's sake we chose to treat these non-detections by considering each upper limit data point as a measured value of zero with its accompanying wide error bar. We checked that our analysis changes very little when upper limit points are omitted. It is nevertheless possible to add the information obtained by these points much more rigorously, as presented in Ref.~\cite{Sawicki_2012,Ishikawa_2022,yung_cdm_2025}.
\begin{table}[h!]
\centering
\small
\begin{tabular}{lcc}
\hline
\textbf{Parameter} & \textbf{Prior type} & \textbf{Range} \\
\hline
\multicolumn{3}{c}{\textbf{Conservative scenario}} \\
\hline
$\epsilon_{\star}$ & Log-uniform or uniform & $[0.001, 1]$ \\
$M_c \space[M_{\odot}]$              & Log-uniform            & $[10^8, 10^{15}]$ \\
$\alpha_{\star}$   & Uniform                & $[-1, 2.5]$ \\
$\beta_{\star}$    & Uniform                & $[-5, 0]$ \\
$G\mu$             & Log-uniform or uniform            & $[10^{-11}, 10^{-6}]$ \\
& & $[10^{-30}, 10^{-6}]$\\
\hline
\multicolumn{3}{c}{\textbf{Fiducial scenario}} \\
\hline
$\alpha_{\star}$   & Uniform                & $[0, 3]$ \\
$\beta_{\star}$    & Uniform                & $[-5, 0]$ \\
$\epsilon_{\star}^{i}$ & Uniform            & $[-3,3]$ \\
$\epsilon_{\star}^{s}$ & Uniform            & $[-3,3]$ \\

$M_c^i $               & Uniform            & $[10^7, 10^{15}]$ \\
$M_c^s $              & Uniform            & $[-3,3]$ \\

$G\mu$             & Log-uniform or uniform            & $[10^{-11}, 10^{-6}]$ \\
& & $[10^{-30}, 10^{-6}]$\\
\hline
\end{tabular}
\caption{Summary of prior choices adopted for the conservative (top) and fiducial (bottom) scenarios in the analysis of UVLFs data from $z=4$ to $z=17$. {Changing the prior intervals for $\alpha_{\star}$ between the conservative and fiducial scenarios does not alter or improve the constraints on the parameters, as shown in Figures \ref{fig:Violin}, \ref{fig:cornerplothst7} and \ref{fig:cornerplotfiducial}.}}
\label{tab:priors_scenarios}
\end{table}
\section{Results}
\label{sec:results}
\begin{figure}
\includegraphics[width=\linewidth]{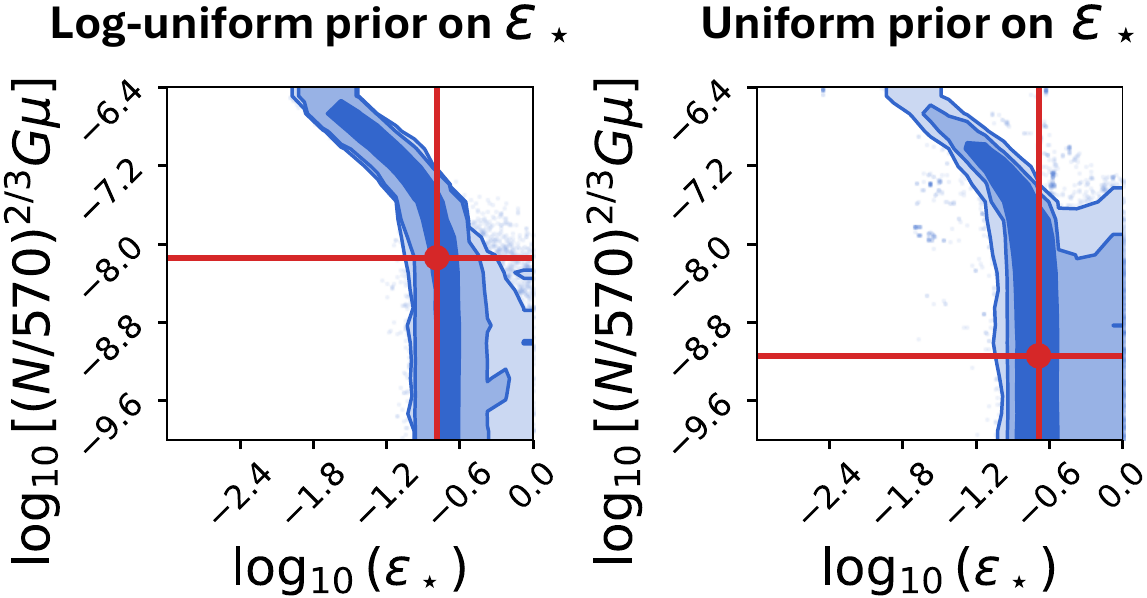}% Here is how to import EPS art
\caption{\label{fig:posthstz7UVLF} Marginalized two-dimensional posterior distribution of $\epsilon_\star$ assuming the conservative scenario at redshift $z=7$ and assuming log-uniform prior on $G\mu$. The color ranges from darkest to lightest correspond to the regions containing $68\%$ percent, $95\%$ percent, and $99\%$ percent of the samples. These results suggest a strong degeneracy between the SFE and the cosmic string tension. Without more information on the SFE, this degeneracy remains unbroken and results in some sensitivity to whether one assumes a log-uniform (left) or uniform (right) prior on $\epsilon_\star$.}
\end{figure}
\subsection{Conservative Scenario}
\subsubsection{HST data}
In the conservative scenario the SFE parameters at different redshifts are completely independent of one another. For redshifts $z=5, 6, 7$, and $8$ (each treated as a separate observation and inference process), we find a strong degeneracy between the SFE and the string tension. As an example, Figure \ref{fig:posthstz7UVLF} illustrates this degeneracy for redshift $z=7$. UVLF observations fix the level of star formation, but are agnostic as to whether the halos hosting the star formation are regular $\Lambda$CDM halos or were seeded from cosmic strings. Therefore, there is a strong degeneracy between the cosmic string tension and the overall normalization of the SFE $\epsilon_\star$. The inclusion of cosmic strings boost star formation, but it is possible to reduce the SFE and to obtain the same UVLF as a scenario with no cosmic strings but a stronger SFE. As shown in Figure \ref{fig:posthstz7UVLF}, for string tensions around $G \mu \gtrsim 10^{-7}$ a low SFE is the only possibility because otherwise one would create too many UV-bright galaxies. However, when $G\mu$ decreases and falls below this value, the cosmic strings play only a minimal or even negligible role and the SFE returns to what is usually seen in $\Lambda \text{CDM}$. The same effect is seen for three of the SFE parameters ($\alpha_{\star}, \epsilon_{\star}, M_c$), while $\beta_{\star}$ is only weakly constrained, as shown in the Figure \ref{fig:cornerplothst7} in Appendix \ref{sec:fitparamvals} presenting posterior constraints.  

The aforementioned degeneracy between $\epsilon_\star$ and $G\mu$ for single-redshift data exists regardless of the choice of priors; however, the precise quantitative limits that can be set will depend on the priors. For example, Figure~\ref{fig:posthstz7UVLF} shows that switching from a log-uniform prior on $\epsilon_{\star}$ to a uniform prior on $\epsilon_{\star}$ limits the degeneracy because this favors higher SFE. As a result, 
one obtains a possible upper bound on $G\mu$ better than is stronger than those from the CMB, since the higher SFE leaves less room for cosmic string-sourced halos. Indeed, a uniform prior on $\epsilon_{\star}$ seems more physically motivated as most studies tend to favor a value of $\epsilon_{\star}$ around $\sim 0.1$ rather than 10 to 100 times smaller \citep{munoz_breaking_2023,sabti_galaxy_2022,sabti_insights_2024,Sabti_2023,yung_cdm_2025,munoz2026relativelyfastreasonablyfurious}.

\begin{figure}[h!]
\includegraphics[width=\linewidth]{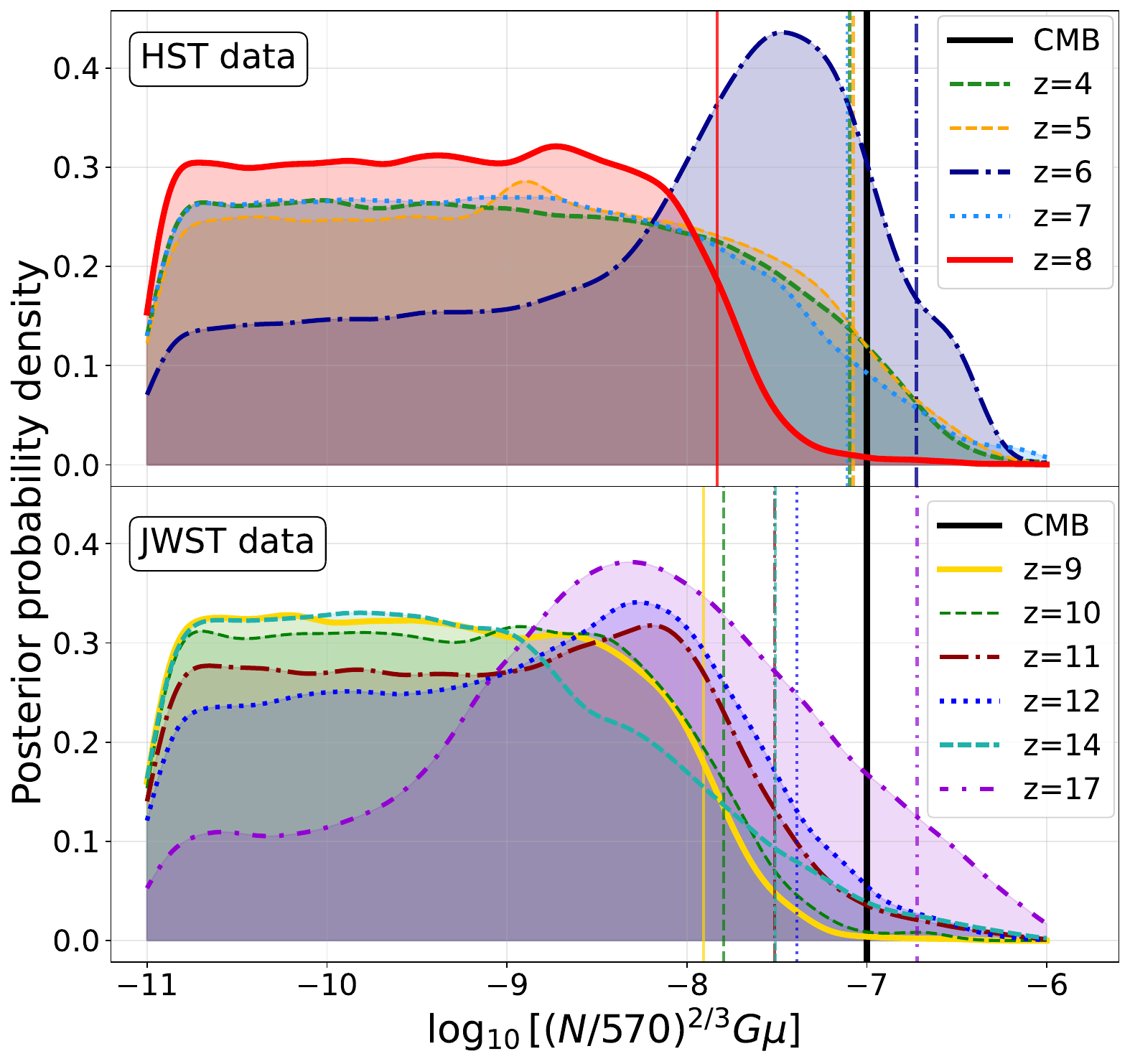}
\caption{\label{fig:densityJWST} Redshift-by-redshift marginal posterior distributions of $\log_{10}[(N/570)^{2/3} G\mu]$ for the conservative scenario with HST UVLFs assuming a log-uniform prior and a prior range on $G\mu$ of $[10^{-11},10^{-6}]$ at redshifts $z=4$ to $z=8$ (top); and with JWST UVLFs for redshifts $z=9$ to $z=17$ (bottom). The vertical lines indicate the upper limits of $\log_{10}[(N/570)^{2/3} G\mu]$ in which $95\%$ percent of the samples are contained contain. The most competitive bound is at redshift 9 and with $N=570$ gives $G\mu\le 1.24\times10^{-8}$. The bounds at $z=11$ and $z=14$ are very similar. }
\end{figure}

The top plot of Figure~\ref{fig:densityJWST} shows the posterior distributions on $\log_{10}[(N/570)^{2/3} G\mu]$ from $z = 4$ to $z=8$ that result from log-uniform priors on $G \mu$ and uniform priors on $\epsilon_\star$. The resulting $95\%$ upper bounds are listed in Table~\ref{tab:results} along with additional information about the priors \footnote{Note that all one-dimensional posterior probability distributions shown in the paper have been smoothed using a kernel density estimator for visualization purposes.}. Most UVLFs (with the exception of redshift $z=6$) seem to rule out cosmic strings with string tension smaller than the CMB limit. The shapes of the posterior distributions are flat and constant for string tensions smaller than $G\mu\le10^{-8}$ and tends to decrease towards larger $G\mu$. 
The strongest constraint on $G\mu$ is obtained at redshift $z=8$ with an upper bound of $G\mu \le 1.47\times10^{-8}$ assuming a prior range on $G\mu$ of $[10^{-11},10^{-6}]$ and $G\mu \le 2.12\times10^{-9}$ assuming a prior range of $[10^{-30},10^{-6}]$.

\subsubsection{JWST data}

Continuing with our conservative scenario (where each redshift analyzed independently with its own set of SFE parameters), we now consider higher redshifts accessible to JWST. Here, the UVLFs tend to allow a wider range of values for the various SFE parameters, close to those of priors with and without cosmic strings. These results are similar to those in Ref. \cite{yung_cdm_2025}. Nevertheless, we will find that with a uniform prior on $\epsilon_{\star}$ and a log-uniform prior on $G\mu$, one still obtains upper bounds on $G\mu$ that are more stringent than that from the CMB.

Degeneracies quite similar to those with the HST data between $\alpha_{\star}$ and $M_c$ have been found, while there is a lack of constraint on $\epsilon_{\star}$. Despite this lack of information on $\epsilon_{\star}$, new constraints can still be placed on $G\mu$. As one goes to high redshifts beyond $z=9$, the importance of the cosmic string halo mass function increases relative to the ST halo mass function (recall Figure~\ref{fig:halomass}). The result is that if $G\mu$ were set to the CMB limit of $G \mu \lesssim 10^{-7}$, we would boost star formation beyond what is consistent with observed high-redshift UVLFs. Figure \ref{fig:densityJWST} (bottom) shows all the marginal distributions on string tension $\log_{10}[(N/570)^{2/3} G\mu]$ obtained.

Figure \ref{fig:densityJWST} shows that while extremely high values of $G\mu$ are disfavored at all redshifts, below $G\mu \sim 10^{-8}$ the constraints roughly split into two classes. At some redshifts (namely $z = 9, 10,$ and $14$) the marginalized posterior distribution rises to a plateau as $G\mu$ decreases. In these cases, the data prefer to boost star formation to explain JWST results. At other redshifts ($z= 11, 12$, and $17$), there is some preference for invoking cosmic strings as an ingredient for explaining UVLFs, and there is a peak around specific string tension values. These peaks are generally consistent with each other but the strongest upper limit of $G\mu\le 1.24\times10^{-8}$ is obtained at redshift $z=9$ assuming a log uniform prior range on $G\mu$ of $[10^{-11},10^{-6}]$. Because the distribution is fairly broad, there is some sensitivity to the lower cutoff of the prior, and $G\mu \le 2.22\times10^{-9}$ is obtained when assuming a prior range of $[10^{-30},10^{-6}]$. For completeness, we also tested for prior sensitivity on $\epsilon_{\star}$, considering a log uniform prior rather than a uniform one. We are confronted with the same degeneracy as presented in the previous section, but in an even more way. Relatively large cosmic string tensions on the order of $\sim 10^{-7}$ become possible if one can tolerate $\epsilon_{\star} \sim 0.002$. While such a value would not be particularly disfavoured by a log uniform prior, it would be (physically speaking) considered an extremely low value. \begin{figure}[]
\includegraphics[width=\linewidth]{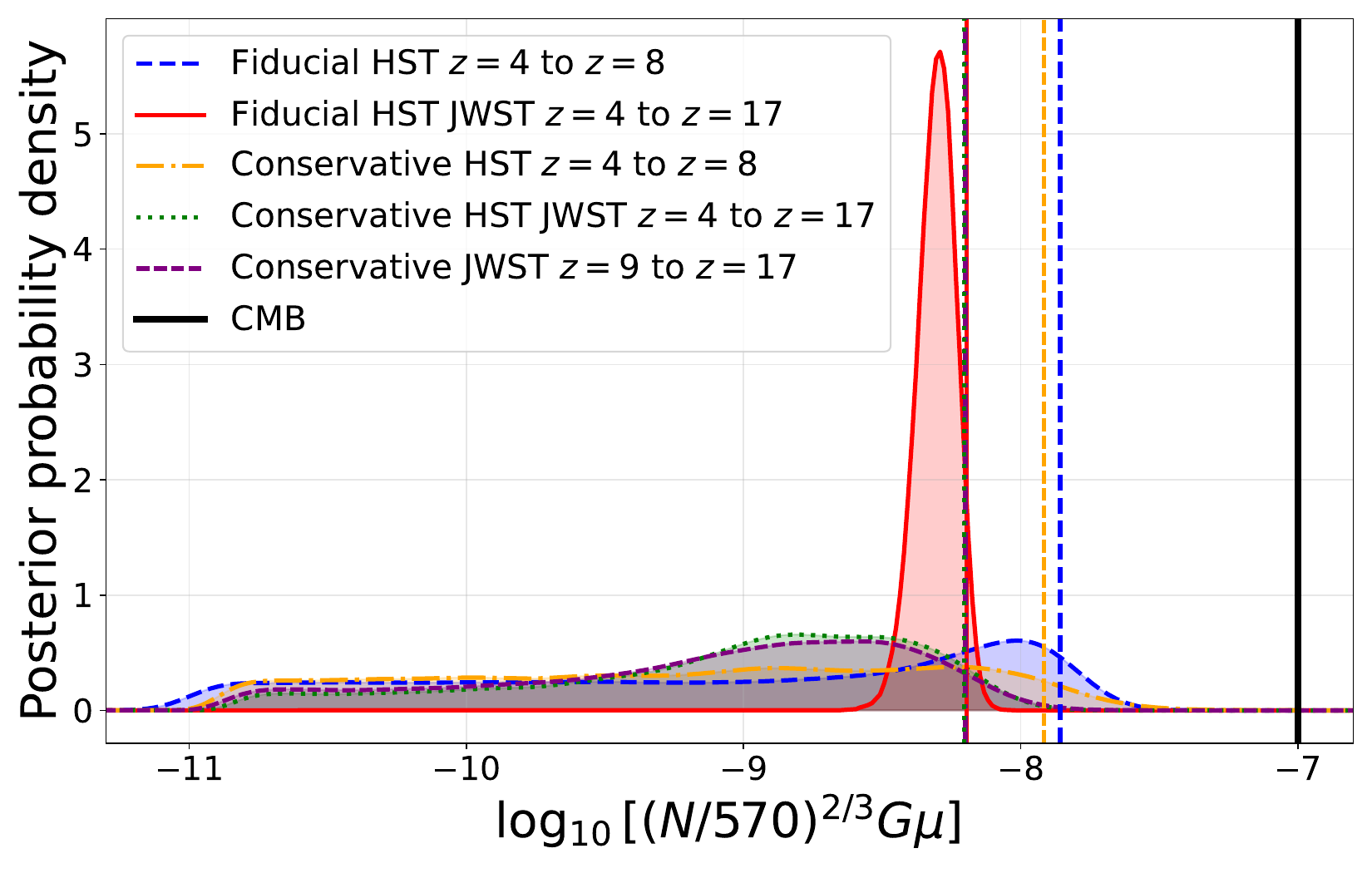}
\caption{\label{fig:density2} Marginal posterior distribution of $\log_{10}[(N/570)^{2/3} G\mu]$ for the fiducial scenario with HST UVLFs alone (blue dashed) at redshifts $z=4,5,6,7$, and $8$ and JWST and HST UVLFs at redshifts from $z=4$ to $17$ (red solid). Also included are the combined constraints from the different redshifts in the conservative case using only HST data (orange dash dotted), HST and JWST data (green dotted), and JWST data only (purple dahsed). The vertical lines indicate the upper limits of $\log_{10}[(N/570)^{2/3} G\mu]$ in which $95\%$ percent of the samples are contained contain.}
\end{figure}
We have also summarized these results in Table \ref{tab:results}.
\begin{figure*}

\includegraphics[width=\linewidth]{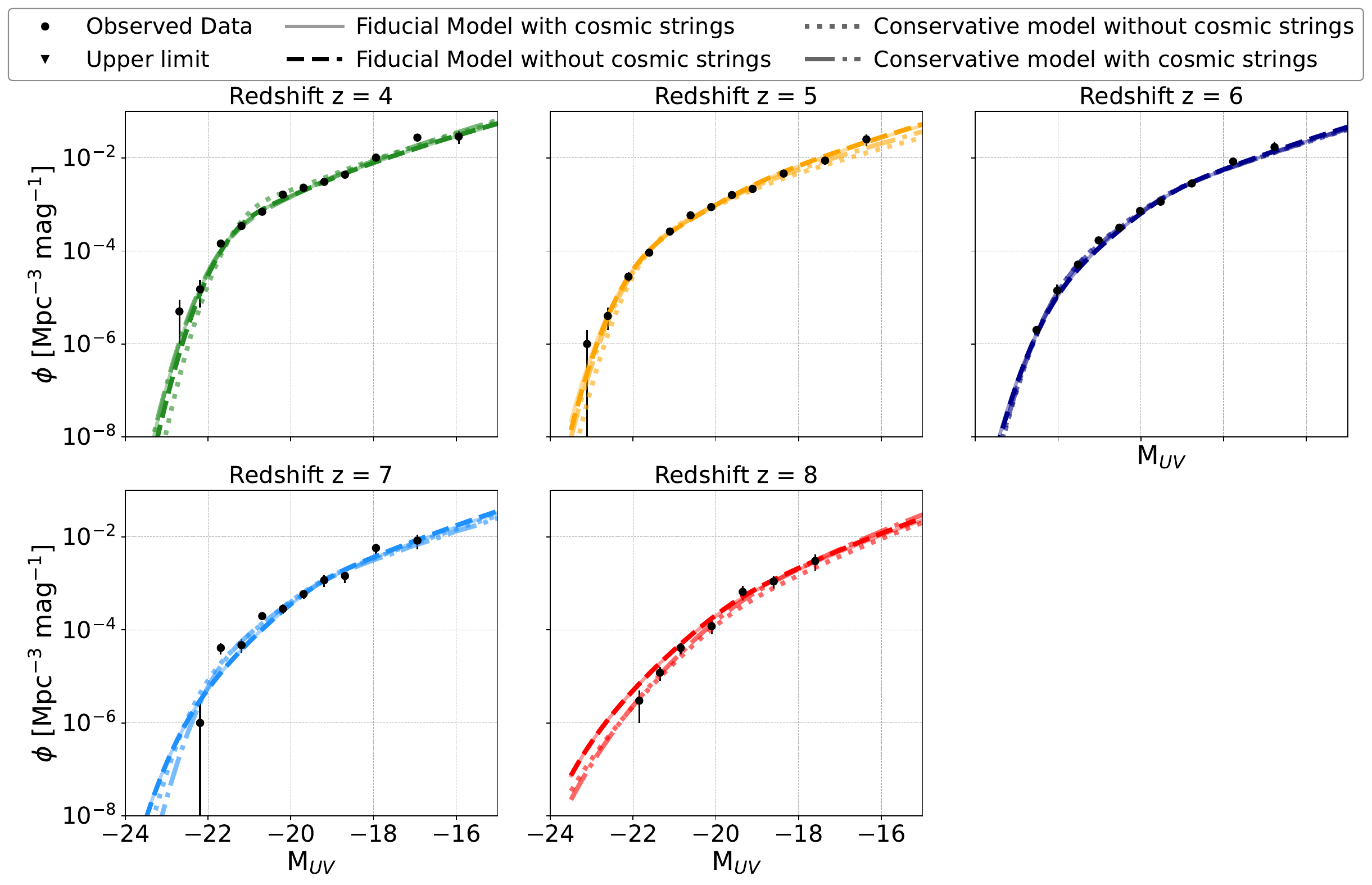}
\caption{\label{fig:fitlowz} Best-fit predictions using the maximum a posteriori estimate obtained from the posterior distribution of parameters assuming fiducial model with (solid) and without (dashed) cosmic strings, and also the conservative model with (solid dashed) and without (dotted) cosmic strings  on HST data from redshifts $z=4$ to $z=8$ described Section~\ref{sec:samples}. The astrophysical parameter values are summarized in Table~\ref{tab:paramsvalues} in the Appendix \ref{sec:fitparamvals}. Since the SFE parameters are degenerate for conservative models, many sets of SFE parameter values yield a good fit with very similar posterior values and we recommend caution when interpreting these SFE parameter values.  At these redshifts, it is not necessary to invoke cosmic strings to get a good fit to the data.}
\end{figure*}

\begin{figure*}
\includegraphics[width=\linewidth]{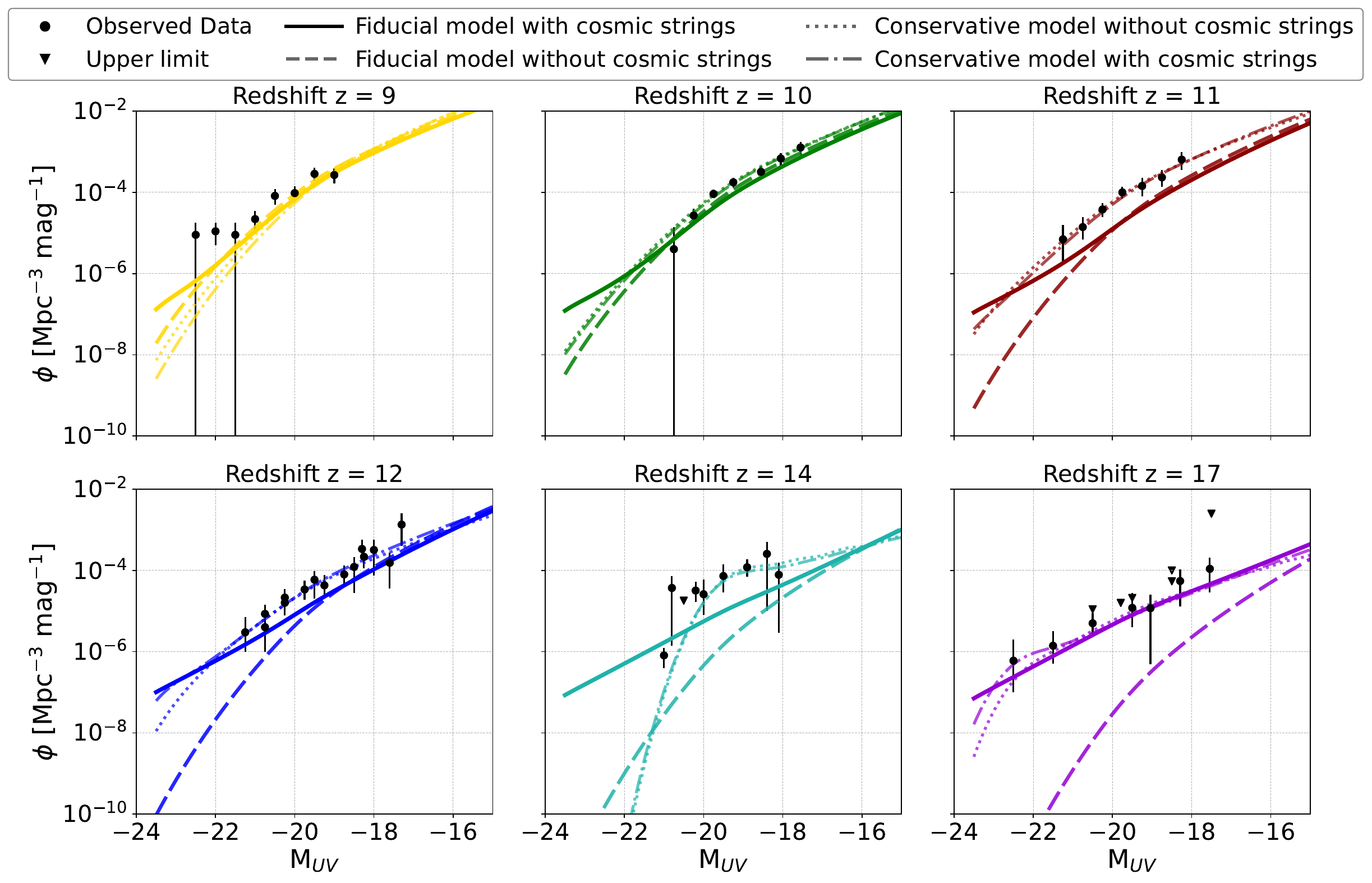}
\caption{\label{fig:fithighz} Same as Figure~\ref{fig:fitlowz} but for JWST data from redshifts $z=9$ to $z=17$ described Section~\ref{sec:samples}. The astrophysical parameters and string tension have the same values as those shown in Figure \ref{fig:fitlowz} and are summarized in Table~\ref{tab:paramsvalues} in the Appendix \ref{sec:fitparamvals}. In the absence of cosmic strings, the SFE with the fiducial model assumption fails to increase sufficiently with redshift, making it impossible to fit the high redshift observations well. In the case where cosmic strings exist, the boost in galaxy abundance produced by cosmic strings makes it possible to compensate for the necessary change in the SFE of the fiducial model and to produce a good fit for the high-redshift data. Once again, we recommend caution when interpreting the SFE parameters obtained in the conservative models, especially since, at these high redshifts, the SFE parameters are even more degenerate. These fits are primarily intended to illustrate that it is possible to explain the abundance of bright galaxies with a more flexible SFE.}
\end{figure*}

\subsection{Fiducial Scenario}

In this scenario, the SFE parameters have a prescribed redshift dependence (as described in Section~\ref{sec:fidscenario}).
% is redshift-dependent and is studied on redshifts $z=4, 5, 6, 7, 8$ observed by HST. 
% The model is in good agreement with the HST data, both with and without cosmic strings. 
This parameterization breaks down the degeneracies between SFE and string tension that we saw when the SFE had the freedom to arbitrarily vary from redshift to redshift. Intuitively, at the lowest redshifts $z = 4$ and $5$, the impact of cosmic strings on UVLFs is negligible. All the data must therefore be explained without recourse to cosmic strings, placing precise constraints on the model parameters in such a way as to give an overall SFE of roughly $\sim 0.2$. With the values of these parameters locked into certain values by the low redshift data, there is no longer any freedom to abruptly change the SFE at high redshift beyond the smooth power law evolution of Equation \ref{eq:fiducialparam}. Within the HST data range of $z \sim 4$ to $8$, this reduces the role that cosmic strings can play without being at odds with the data, since the vanilla star formation model alone is (somewhat by construction) in consistent with all HST data. This can be seen in Figure~\ref{fig:fitlowz}, where one sees that the data are equally well-fit with and without cosmic strings. We can also observe a weaker degeneracy between SFE parameters and $G\mu$, as seen in the constrained posterior distributions in Figure \ref{fig:cornerplotfiducial} of the Appendix \ref{sec:fitparamvals}. Figure \ref{fig:density2} (blue) shows the marginal posterior distribution of $\log_{10}[(N/570)^{2/3} G\mu]$ for $4 < z < 8$. With this parametrization and assuming $N=570$, the upper bound obtained on $G\mu$ is $G\mu\le 1.9\times10^{-8}$.

% \subsection{A possible solution to explain both HST and JWST data}

The situation is different when JWST-relevant redshifts are included, because the vanilla string-less scenarios do not easily fit the UVLF data. Indeed, in previous studies the SFE tends to jump from $\sim 0.1$ to $\sim 0.4$ at $z \gtrsim 9$ if it is allowed to vary freely from redshift to redshift. The redshift-dependent parameterization does not do much better, simply because the chosen parameterization does not allow the SFE to grow quickly enough at the higher redshifts. In other words, the data demand a new ingredient.

Cosmic strings can potentially be this ingredient, and our model is able to achieve good agreement with data at most redshifts. This is illustrated by the red curve in Figure~\ref{fig:density2}, which shows the posterior distribution considering data from the entire range in redshift. Surprisingly, the posterior distribution seems naively to suggest evidence of cosmic strings, not only for a very precise value of $G\mu=5.08\times 10^{-9} \pm 1.58\times 10^{-9}$ but also for the SFE parameters. Naively, this seems to suggest evidence of cosmic strings with a very precise value of $G\mu=( 5.08 \pm 1.58) \times 10^{-9} $. The SFE parameters also favour fairly precise values. Intuitively, what happens is that going from low redshifts to high redshifts, the SFE decreases slowly at first, but eventually the string-seeded halos take over as a non-negligible contribution to star formation at high redshifts. Our model is able to fit the high-redshift data quite well, as shown in Figure~\ref{fig:fithighz}.

Our rather precise constraint on $G\mu$ can be qualitatively understood in two ways. First, it is an illustration of cosmic string physics. At low redshift, cosmic strings have a negligible contribution, while at higher redshift, they will boost halo formation and be consistent with the data. Since JWST data demand a particular boost in the UVLFs and our model delivers this through cosmic strings, a very precise value of $G\mu$ is implied. Second, this precise value is a reflection of the radical change in UVLFs when analyzing JWST's redshift data. If the JWST data had evolved smoothly from HST as a function of redshift, our analysis would provide only an upper bound on $G\mu$ and not an upper bound and a lower bound. By keeping an SFE smooth and parameterized on HST data, the existence of a lower bound on $G\mu$ is simply suggestive of a need to somehow boost our UVLFs. On the other hand, the upper bound indicates that we should not boost halo formation too much at the risk of overshooting and predicting UVLFs that are too high in amplitude. 

In Figure \ref{fig:density2}, we have also added for the sake of comparison the combined constraints from the conservative scenario---essentially the multiplication of the curves from Figure~\ref{fig:densityJWST}---from redshifts $z = 4$ to $z=8$ (orange dashed dotted), from redshifts $z=9$ to $z=17$ in (purple dashed), and $z=4$ to $z=17$ (green dotted). By comparing the marginal posterior distributions in Figure \ref{fig:density2} between the conservative and fiducial scenarios, we can directly see that the seemingly strong evidence for cosmic strings disappears and is really specific to the choice of SFE parameterization. The peaks, when present, are now much broader and in some cases disappear entirely into a plateau. That said, one sees that in all of these scenarios new upper bounds on $G\mu$ can be obtained than are more stringent than those from the CMB. The exact bounds are summarized in Table~\ref{tab:results}.

% To confirm the parameterization dependence, we have also included Figure \ref{fig:Violin}, which shows the possible solution of cosmic strings in broken dashed lines compared with the posterior distributions of the conservative scenario for all parameters as a function of redshift. In the conservative scenario at high redshift, we can see that the uncertainties on the SFE are very large and there is no strong evidence for a specific string tension. In fact, the cosmic string solution is only viable for a very precise string tension and for very precise SFE parameters, in other words, a specific parameterization, the fiducial scenario with very specific SFE values. These arguments confirm that cosmic strings are a viable solution for explaining the abundance of bright galaxies, but only when keeping the star-formation physics linearly evolving with redshift.

\begin{figure}[]
\includegraphics[width=\linewidth]{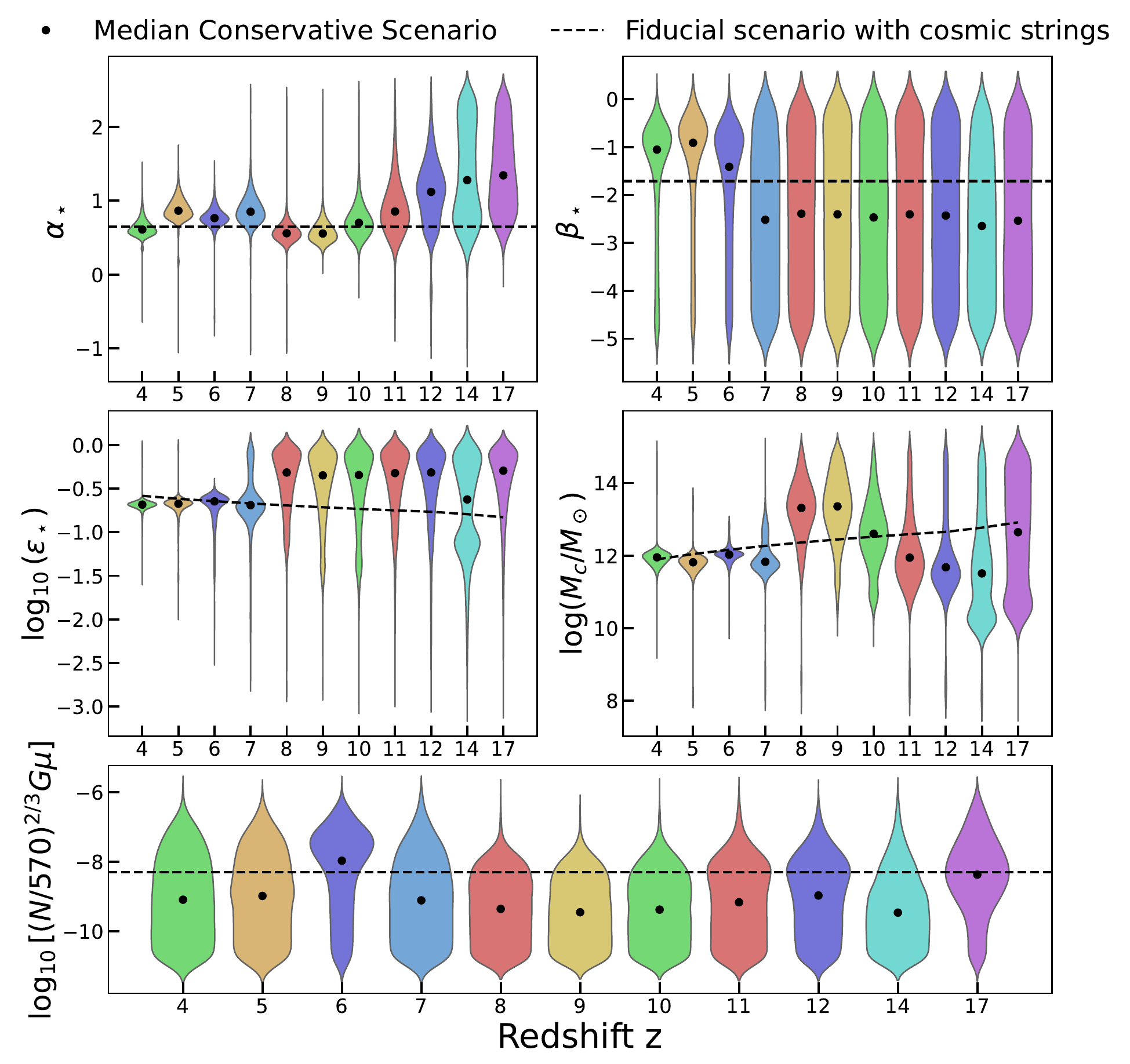}% Here is how to import EPS art
\caption{\label{fig:Violin} Comparison of the conservative (violin) and fiducial scenarios with cosmic strings (dashed line) for the four SFE parameters and $G\mu$ as a function of redshift. For the fiducial scenario, error bars are not included but are of the order of 10 percent for the non-constant SFE-related parameters and a few percent for the rest, including $G\mu$. The parameter $\beta_{\star}$ remains unconstrained. While the fiducial scenario with cosmic strings allows for a correct fit of all HST and JWST data, the conservative scenario suggests that the SFE remains largely unconstrained.}
\end{figure}

To confirm the parameterization dependence, consider Figure \ref{fig:Violin}, which shows a possible solution of cosmic strings from the parameterized fiducial scenario in broken dashed lines compared with the posterior distributions of the conservative scenario for all parameters as a function of redshift. In the conservative scenario at high redshift, we can see that the uncertainties on the SFE are very large and there is no strong evidence for a specific string tension. In fact, the cosmic string solution is only viable for a very precise string tension and for very precise SFE parameters. In other words, it is suggestive of a specific evolution that can be encapsulated in a parametrization of redshift evolution. These arguments confirm that cosmic strings are a viable solution for explaining the abundance of bright galaxies, but only when tying together the star-formation physics smoothly from redshift-to-redshift as we do in the fiducial scenario.

\begin{table*}[t]
\centering

\setlength{\tabcolsep}{6pt}
\renewcommand{\arraystretch}{1.35}

\begin{tabularx}{\textwidth}{l c c X}
\toprule
\textbf{Scenario} & 
\textbf{Upper bound on $G\mu$} &
\textbf{Upper bound on $G\mu$} &
\textbf{Prior dependence} \\
\midrule

\multirow{1}{*}{\textbf{Conservative Scenario}}
    & Log-uniform prior on $G\mu$
    & Log-uniform prior on $G\mu$
    & \multirow{13}{\hsize}{Bounds on $G\mu$ are sensitive to the prior choice and lower bound on the prior. Quoted limits assume a log-uniform prior on $G\mu$ and uniform prior on $\epsilon_\star$. } \\
    &$[10^{-11},10^{-6}]$&$[10^{-30},10^{-6}]$&\\
   HST $z = 4$ & $8.02\times 10^{-8}$ & $6.00\times 10^{-9}$ & \\
   HST $z = 5$ & $8.39\times 10^{-8}$ & $1.37\times 10^{-8}$ & \\
    HST $z = 6$& $1.89\times 10^{-8}$ & $5.96\times 10^{-8}$ & \\
    HST $z = 7$& $7.74\times 10^{-7}$ & $2.59\times 10^{-9}$ & \\
    HST $z = 8$& $1.47\times 10^{-8}$ & $2.11\times 10^{-9}$ & \\
    JWST $z = 9$& $1.24\times 10^{-8}$ & $2.22\times 10^{-9}$ & \\
    JWST $z = 10$ & $1.60\times 10^{-8}$ & $2.53\times 10^{-9}$ & \\
    JWST $z = 11$& $3.08\times 10^{-8}$ & $5.67\times 10^{-9}$ & \\
    JWST $z = 12$& $4.07\times 10^{-8}$ & $1.55\times 10^{-8}$ & \\
    JWST $z = 14$& $3.10\times 10^{-8}$ & $8.50\times 10^{-10}$ & \\
    JWST $z = 17$& $1.90\times 10^{-7}$ & $5.00\times 10^{-8}$ & \\
\midrule

\multirow{1}{*}{\textbf{Fiducial scenario}} 
&Log-uniform prior on $G\mu$ & Uniform prior on $G\mu$&\multirow{3}{\hsize}{No sensitivity to the choice of uniform or log-uniform priors and no sensitivity of the choice of lower bound of the prior.}\\
 &$[10^{-11},10^{-6}]$ & $[10^{-11},10^{-6}]$  \\
HST $z=4$ to $z=8$  & $1.39\times 10^{-8}$ & $1.85\times 10^{-8}$ & \\
 HST JWST  $z=4$ to $z=17$     & $6.37\times 10^{-9}$ & $6.37\times 10^{-9}$ & \\
   
\midrule

\multirow{1}{*}{\textbf{Combined posterior distribution}} 
    &  Log-uniform prior on $G\mu$
    & Log-uniform prior on $G\mu$
    & \multirow{5}{\hsize}{Slightly sensitive to the choice of uniform or log-uniform priors and slightly sensitive of the choice of lower bound of the prior (sensitive for HST)} \\
    &$[10^{-11},10^{-6}]$&$[10^{-30},10^{-6}]$&\\
    HST $z=4$ to $z=8$& $5.07\times 10^{-9}$ & $1.61\times 10^{-8}$ &\\
    JWST $z=9$ to $z=17$& $3.73\times 10^{-9}$ & $8.11\times 10^{-9}$ & \\
    HST JWST $z=4$ to $z=17$& $2.47\times 10^{-9}$ & $8.21\times 10^{-9}$   \\
\bottomrule

\end{tabularx}

\caption{\label{tab:results} Summary of the new $95\%$ upper bounds obtained on $G \mu$ for a variety of prior assumptions and supposing $N=570$.}

\end{table*}

\section{Discussion}
\label{sec:discussion}

\subsection{Are the Sheth $\&$ Tormen and the cosmic string loop halo mass functions valid?}

A key quantity that we have relied on in this paper is the halo mass function, which we have assumed to take the ST form and then perturbed in the presence of cosmic strings. Although the ST form has been largely confirmed by various numerical simulations, recent simulations specially designed to obtain the HMF at high reshifts have shown a possible mismatch between the ST halo mass function and that obtained in GUREFT simulations \citep{yung_cdm_2025,yung2024GU,yung_are_2024}. The halo mass function obtained by GUREFT simulations is higher than ST in the low-mass ($\lesssim 10^{10}M_\odot$) regime, but exponentially lower than ST in the high-mass ($\gtrsim 10^{10}M_\odot$) regime. From Figure~\ref{fig:halomass}, we know that the latter mass range is where the relative influence of cosmic strings is greatest. Therefore, if we had used the GUREFT halo mass function instead of ST, the upper limits on $G\mu$ would likely have been more conservative.

There are also uncertainties surrounding the HMF of cosmic strings. Our study was based on the HMF derived in Ref.~\cite{jiao_early_2023}, which assumed that cosmic string loops have negligible velocity and are static. However, this static assumption was challenged by Ref.~\cite{cyr_not-quite-primordial_2025} which included the velocities of these strings and the resultant modifications to their accretion history. While some uncertainties remain regarding the exact impact of velocities, it is likely that there would be some effect on $G\mu$ constraints. In Appendix~\ref{sec:velocityappendix}, we implement the halo mass function of Ref.~\cite{cyr_not-quite-primordial_2025} and examine the changes to our cosmic string constraints.

Given the caveats presented here, it is clear that a definitive study of cosmic strings using UVLFs (beyond the proof-of-concept-level constraints of this paper) will require the reduction of uncertainties in halo mass function modeling, whether for standard halos or cosmic string-seeded halos.

\subsection{On the choice of priors for $\epsilon_\star$ and $G\mu$}

As briefly discussed in Section~\ref{sec:results}, at any \emph{individual} redshift, with UVLFs alone it is difficult to distinguish between a scenario with a larger number of dark matter haloes combined with a very low SFE and the opposite case with fewer dark matter haloes and a higher SFE. As such, for single-redshift constraints the choice of prior on $G\mu$ and $\epsilon_\star$ become quite important. The degeneracy is particularly severe when both parameters receive uniform priors or both parameters receive log-uniform priors. In those scenarios it is easy for there to be a high string tension paired with a very low SFE. The SFEs that result are generally one to two orders of magnitude lower than expected in standard analyses, and are therefore arguably not very physical. We have nonetheless performed these analyses as an exploration, but consider it better motivated to use a log-uniform prior on $G \mu$ and a uniform prior on $\epsilon_\star$ to favor more physical SFEs.

It is also important to note that in this conservative redshift-by-redshift scenario, the $95\%$ upper bound on $G\mu$ depends on the lower bound of the log-uniform prior on $G\mu$. Reducing the lower bound shifts the final results toward a smaller value of $G\mu$. This effect is related to the shape of the posterior marginal distribution, which is plateau-shaped and then drops. For this reason, we have also included a study of posterior distributions with the broadest prior interval on $G\mu$ of $[10^{-30},10^{-6}]$ where the $10^{-30}$ lower limit is set by the energy scale of Large Hadron Collider. Table~\ref{tab:results} summarizes the variation that results from different lower bounds.

When we combine the redshift-by-redshift constraints on $G\mu$ by multiplying together their likelihoods, we find that there is a slight reduction in sensitive to prior choice. This is mostly driven by the slightly peaked nature of the $z=6,11,12,$ and $14$ constraints in Figure~\ref{fig:densityJWST}. However, there is still some sensitivity to both the form of the prior as well as the choice of the lower bound, as one can see from Table~\ref{tab:results}.

In contrast, for the fiducial scenario with the parameterized redshift dependence on SFE, the prior sensitivity largely vanishes. From Table~\ref{tab:results}, we see that this is especially true when jointly fitting both HST and JWST data. This does not mean, however, that one need not be concerned about priors. In essence, the choice of parameterization itself is a type of prior, reflecting our previous assumptions about star formation physics.

\subsection{Clustering to break degeneracies}
As pointed out by Ref.~\cite{munoz_breaking_2023}, there are several ways of obtaining the UVLFs with different halo-galaxy connections. For example, a large stochasticity in the formation of luminous objects is intrinsically degenerate with the SFE. This is also the case in our analysis where a larger presence of massive halos is intrinsically degenerate with the SFE. A possible approach to try to break this degeneracy will be to look at the clustering of galaxies. In general, galaxies formed in the most massive halos will be strongly clustered than those formed in less massive halos. This clustering can be described and measured by the galaxy bias. In relation to this study, if cosmic strings tend to boost the formation of dark haloes, we can expect a prediction of cosmic strings on the galaxy bias. Such predictions of the galaxy bias from cosmic strings as a function of redshift would also help to break down the degeneracies between star formation and cosmic strings. This will likely be of near-term relevance given the possibility of existing and future experiments measuring galaxy biases up to redshift $z\gtrsim 10$ \cite{Harikane_2022,La_Plante_2023,Waters_2016}. 
\subsection{Future UVLFs and other galaxy observations and  will help constrain cosmic strings}
JWST has opened up a new window on the observation of galaxies at high redshift, with potential detections up to redshift $z \sim 25$ \citep{perez-gonzalez_rise_2025,castellano_pushing_2025}. 
%Although, for the moment, observations at ultra high redshift are mainly photometric, they are tending to be confirmed by spectrocopic observations.  
Although these photometric measurements may be interlopers at lower redshift, they can nevertheless be used as robust upper bounds for UVLFs at ultra high redshifts. More precise data, even in the form of upper bounds, will be extremely useful for constraining cosmic strings. Part of this stems from the consistency across redshifts, since a given string tension will have to be in agreement with UVLF measurements at all redshifts. In just a few years, the JWST has delivered some impressive results, and future observations will make it possible to probe a wider range of mass scales via $M_{UV}$ and potentially higher redshifts. This new observation window also complements others including gravitational waves, CMB, and intensity mapping \citep{Brandenberger_2010,Pagano_2012,Th_riault_2021,Maibach_2021,Hern_ndez_2012,McDonough_2013,Blamart_2022,Brandenberger_2014,bernardo2023superfluiddarkmatterflow,Brandenberger_2019,brandenberger2013angularpowerspectrumbmode,DANOS_2010,Ciuca_2019,global}. 

In fact, one of the main obstacles to better upper bounds on $G\mu$ at high redshift is the lack of understanding of the SFE at ultra-high redshift. For example, in Figure \ref{fig:Violin}, the posterior distribution is extremely wide for most SFE parameters. A better understanding of this SFE will directly help to extract more information about the cosmic string. The wide spreads on the SFE parameters, especially at high redshifts, currently make it difficult to draw precise inferences about $G\mu$.

\section{Conclusion}
\label{sec:conclusions}
In this paper, we first presented UVLFs as a new window for studying cosmic strings. This enabled us to compare cosmic string constraints with UVLF data measured by HST and JWST. This led to a possible explanation and reconciliation of HST and JWST data without the need for abrupt changes in SFE or high stochasticity at high redshifts. It also led to a new upper limit $G\mu\lesssim10^{-8}$ on cosmic string tension representing a factor of ten improvement over the \textit{Planck} 2014 upper limit of $G\mu\le10^{-7}$. To do this, we modified the semi-analytical \texttt{Zeus21} code in order to obtain rapid predictions for a whole set of astrophysical parameters, but also to understand and take into account possible degeneracies between different astrophysical effects and cosmic string effects. We have noted that the current uncertainties on the SFE, particularly on JWST data, are a major limitation in obtaining better upper bounds on cosmic strings. From a general point of view, it is also interesting to note that a high redshift HMF that decreases as a power law in mass (as with cosmic string), rather than exponentially as in the ST HMF, may be considered a plausible solution to explain UV bright galaxies, on condition that it recovers its exponential decrease at lower redshifts.  A better future understanding of the astrophysical properties of these galaxies with precise measurements of the SFE, as well as a better understanding of the formation of dark matter haloes and galaxies seeded by cosmic strings, especially in the regime of high masses and large loops, will greatly improve the upper bounds or potentially provide real evidence of cosmic strings.

\begin{acknowledgments}
The authors are delighted to thank Jiao Hao for help with implementing the halo mass function of cosmic strings and to thank Laurie Amen,
 Marek Detière-Venkatesh, Hannah Fronenberg, Josh Goodeve, Patrick Horlaville, Andrei Li, Kim Morel, Robert Pascua, Sophia Rubens, Debanjan Sarkar, Michael
Wilensky for their support, interesting discussions and valuable advice in the realisation of
this project. MB would like to thank Jonathan Pober and Brown University for their hospitality. 
AL acknowledges support from the Natural
Sciences and Engineering Research Council of Canada
through their Discovery Grants Program and their Alliance International Program, as well as
the William Dawson Scholar program at McGill University. RB is supported in part by a NSERC Discovery Grant and by the Canada Research Chair program. JBM acknowledges support from NSF Grants AST-2307354 and AST-2408637, and the CosmicAI institute AST-2421782. This research was supported in part by grant NSF PHY-2309135 to the Kavli Institute for Theoretical Physics (KITP). BC is grateful for support from an NSERC Banting Fellowship, as well as the Simons Foundation (Grant Number 929255). This research was enabled in part by support provided by 
\href{https://www.calculquebec.ca/}{Calcul Quebec} 
and the \href{https://www.alliancecan.ca/}{Digital Research Alliance of Canada}.
\end{acknowledgments}

\appendix

\section{Inclusion of cosmic string velocities in the analysis} 
\label{sec:velocityappendix}
Towards the end of this project, Ref.~\cite{cyr_not-quite-primordial_2025} challenged the hypothesis of static accretion of cosmic string loops by including the velocity distribution of the cosmic string loop network. Based on numerical simulations, the velocity distribution of the strings can be written as \begin{equation}
f(v_{\text{form}})=Bv_{\text{form}}^2(1-v_{\text{form}}^2)^p,
\end{equation}
with $v_{\text{form}}$ the speed of a loop when it forms. Numerically, $p$ tends to be $p = 10$ and $\langle v_{\text{form}}\rangle \approx 0.3$ in natural units. This
leads to a normalisation constant of $B \approx 85$.

\begin{figure}
\includegraphics[width=\linewidth]{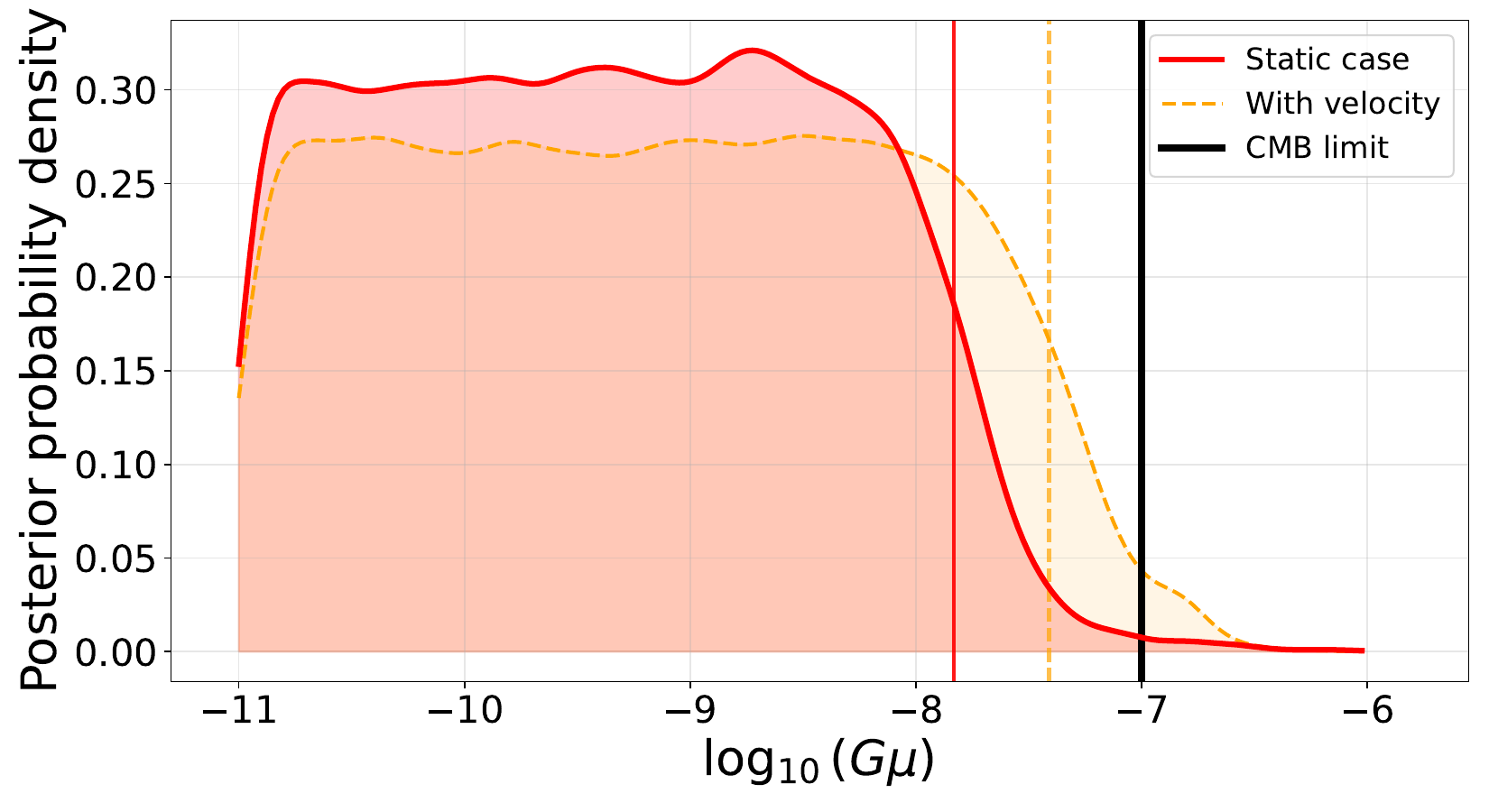}
\caption{\label{fig:densitynewhmf} Comparison of the marginal posterior distribution on $\log_{10}(G\mu)$ for the conservative scenario HST UVLF at redshift $z=8$. The red posterior distribution corresponds to the static HMF similar to that in Figure~\ref{fig:density2} and the yellow/orange to the HMF including velocity effects. Taking into account the speeds of the strings, the limits on string tension are relaxed and tend toward a higher value but remain more stringent than those derived from the CMB.}
\end{figure}

 Depending on the speed of the loop, three different accretion scenarios will come into play and modify the halo mass function. The first, where the speed of the strings is negligible, corresponds to the static case but is subdominant. The second case, with a non-negligible velocity, corresponds to cylindrical accretion known as filamentary accretion. The last scenario leads to fragmentation in the formation of the structure due to the very high velocity of the loops. 
This introduces a modification to a new HMF $dn_{\text{CS}}/dM_h$ written as 
\begin{equation}
\dfrac{dn_{\text{CS}}}{dM_h}=\dfrac{dn_{\text{static}}}{dM_h}+\dfrac{dn_{\text{fil}}}{dM_h}+ \dfrac{dn_{\text{frag}}}{dM_h}
\end{equation}
with $dn_{\text{static}}/dM_h$ corresponding to the HMF where the speed of the strings is negligible, $dn_{\text{fil}}/dM_h$ the HMF contribution from filamentary accretion, and $dn_{\text{frag}}/dM_h$ corresponding to the fragmented accretion.
The biggest change is that the dominant term, $dn_{\text{frag}}/dM_h$, varies as $M_h^{-8/3}$ and with redshift as $(1+z)^{-2}$ rather than $M_h^{-5/2}$ and $(1+z)^{-3/2}$ in the static case discussed above (Equation \ref{eq:halomassloop}). This leads to a decrease in the production of dark matter haloes, particularly in the high-mass regime. 
We repeated the analysis for cases where the upper bounds on $G\mu$ are strongest, i.e., for UVLFs at redshift $z=8$ from HST and $z=9$ from JWST in conservative cases, but also in the combined HST and HST-JWST fiducial scenarios. 
In terms of results, this leads to a less stringent  upper bound on $G\mu$ in all analyses, but nevertheless remains an improvement over the Planck upper bound as presented in Figure \ref{fig:densitynewhmf}. This lead to an upper bound of $G\mu\le 4.36\times10^{-8}$ for $z=8$ and of $G\mu\le 3.90\times10^{-8}$ for $z=9$ with a prior on $G\mu$ of $[10^{-11},10^{-6}]$.

These limits should, however, be treated with caution. As discussed in Ref. \cite{cyr_not-quite-primordial_2025}, this new mass function is only valid between a mass range $[M_{min}, M_{max}]$. The minimum limit corresponds to the fact that, above this mass, the loops corresponding to this mass will also decay and therefore the accretion by these loops will be strongly impacted by this decay. For this analysis, the halos originating from these loops have very little chance of hosting galaxies and therefore make a completely negligible contribution to the HMF. 
For the upper limit of the mass range, this corresponds to the fact that, above this mass, the oscillation effects of the loop come into play. These effects are not taken into account, and thus a high-mass cutoff is imposed to enforce the regime of validity of the present treatment. Unfortunately, the mass range of this high-mass cutoff is relevant to UVLFs in the fiducial scenario at ultra-high redshifts, and the presence of a cutoff reduces the goodness of fit to the data. 
We therefore also tested the analysis in which we extended the new HMF by extrapolating the new halo mass function in the high-mass regime. This allows all UVLFs to be well fit by the model with results similar to those shown in Figure \ref{fig:fitlowz} and Figure \ref{fig:fithighz}. The final limit with this treatment is $G\mu=2.05\times10^{-8}$.

With these variations in results, we therefore emphasize the importance of accurately understanding the accretion of matter around large cosmic string loops for future UVLFs constraints on cosmic strings. \vspace{0.3cm}

\section{Fit parameter values and posterior constraints}
\label{sec:fitparamvals}

In this section, we present Table \ref{tab:paramsvalues} summarizing the values of the astrophysical parameters and $G\mu$ for the UVLF plots in Figure \ref{fig:fitlowz} and Figure \ref{fig:fithighz}. These parameters are obtained using the maximum a posteriori estimate obtained from the posterior distribution of parameters for the conservative and fiducial models. As described in the captions for Figures \ref{fig:fitlowz} and \ref{fig:fithighz}, many SFE parameter values yield a good fit with very similar posterior values, and we recommend caution when interpreting these values.

\begin{figure*}
\includegraphics[width=\linewidth]{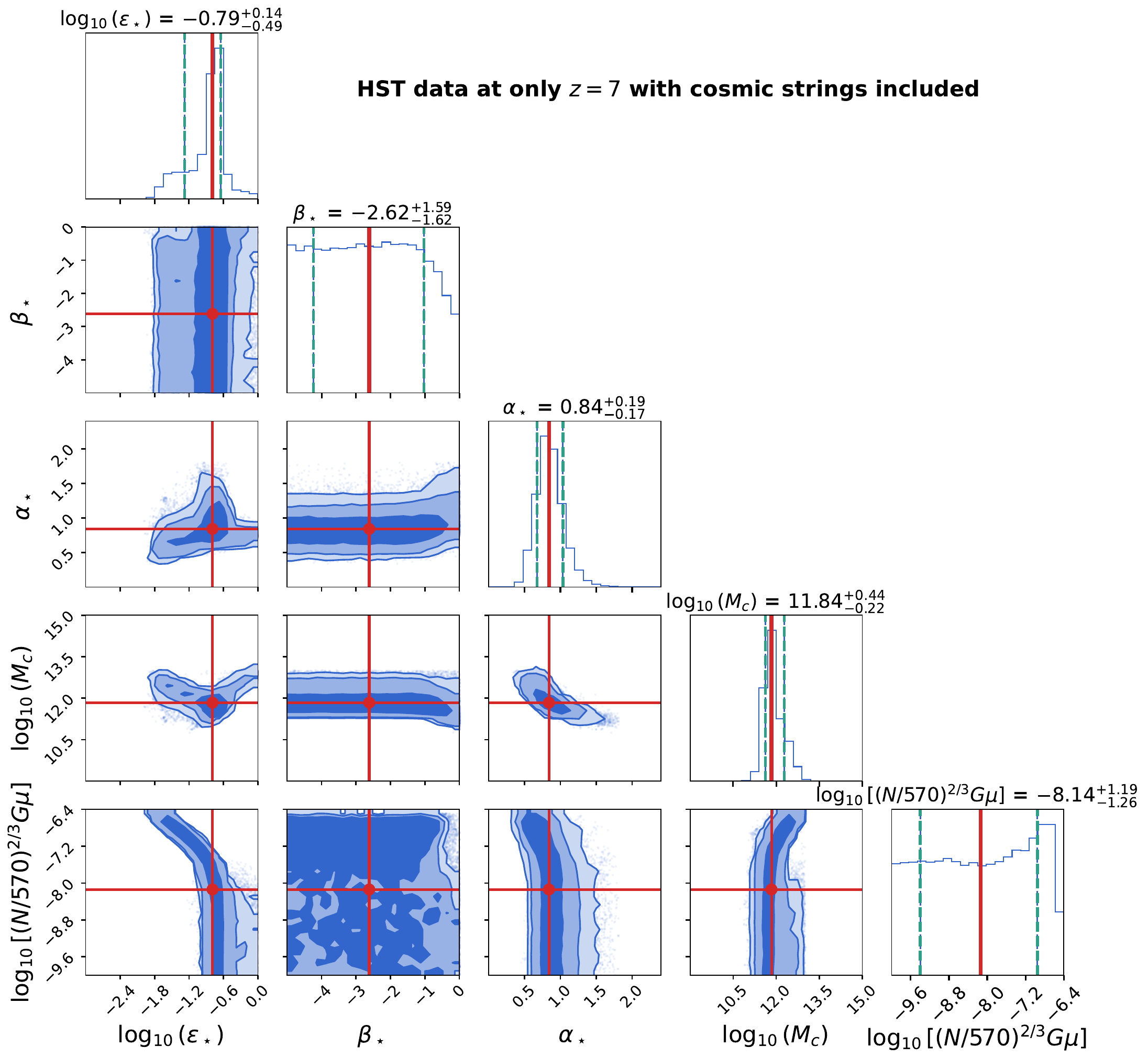}
\caption{\label{fig:cornerplothst7}Posterior constraints on the five parameters of the conservative model Equation \ref{eq:conservativeparam} based on HST data described in Section~\ref{sec:samples} at redshift $z=7$ only. The color ranges from darkest to lightest correspond to the regions containing $68\%$ percent, $95\%$ percent, and $99\%$ percent of the samples. These results assumes a log-uniform on $\epsilon_\star$.}
\end{figure*}

\begin{figure*}
\includegraphics[width=\linewidth]{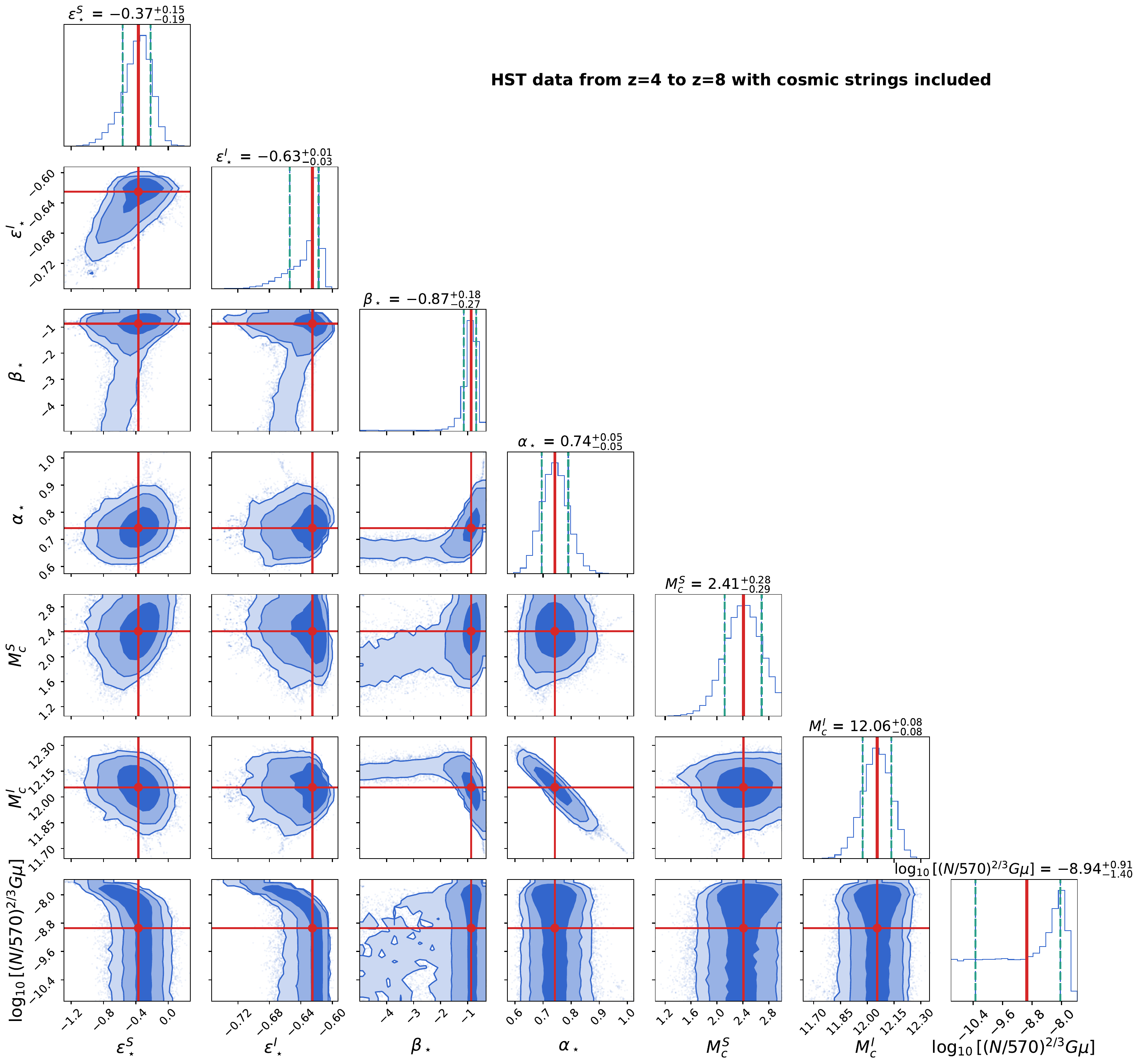}
\caption{\label{fig:cornerplotfiducial} Posterior constraints on the seven parameters of the fiducial model in Equation \ref{eq:fiducialparam} based on HST data described in Section~\ref{sec:samples} from redshift $z=4$ to $z=8$. The color ranges from darkest to lightest correspond to the regions containing $68\%$ percent, $95\%$ percent, and $99\%$ percent of the samples.}
\end{figure*}

In addition, we have also included in Figure \ref{fig:Violin} the violin plot showing the constraints obtained and the evolution on $G\mu$ and the SFE parameters as a function of redshift in the conservative scenario and for the fiducial scenario respectively. In the same vein, to fully understand and present the degeneracies that arise between the SFE and $G\mu$, we present in detail the posterior constraints obtained on the parameters of the SFE and $G\mu$ from the HST data at redshift $z=7$ for the conservative model in Figure \ref{fig:cornerplothst7}, as well as those obtained from the HST data at redshifts $z=4$ to $z=8$ for the fiducial model in Figure \ref{fig:cornerplotfiducial}.

\begin{table}[h!]
\centering
\renewcommand{\arraystretch}{1.25}
\resizebox{0.5\textwidth}{!}{%
\begin{tabular}{lccccccccc}
\toprule
\textbf{Fiducial model} & \multicolumn{9}{c}{\textbf{Parameters}} \\
\cmidrule(lr){2-10}
 & $\alpha_\star$ & $\beta_\star$ & $\epsilon_{\star}^s$ & $\epsilon_{\star}^i$ & $M_c^s$ & $M_c^i$ & \multicolumn{3}{c}{$G\mu$} \\
\midrule
\textbf{Without CS} & 0.69 & $-1.20$ & 0.20  & $-$0.62 & 1.62 & 12.12 & \multicolumn{3}{c}{0} \\
\textbf{With CS}    & 0.65 & $-$1.71 & $-$0.45 & $-$0.65 & 1.85 & 12.16 & \multicolumn{3}{c}{$5.08\times10^{-9}$} \\
\toprule
\textbf{Conservative model} & \multicolumn{9}{c}{\textbf{Parameters}} \\
\textbf{Without CS} & \multicolumn{9}{c}{} \\
\cmidrule(lr){2-10}
 & \multicolumn{2}{c}{$\alpha_\star$} & \multicolumn{2}{c}{$\beta_\star$} & \multicolumn{2}{c}{$\epsilon_{\star}$} & \multicolumn{2}{c}{$\log_{10}(M_c)$} & \\
\midrule
$z = 4$  & \multicolumn{2}{c}{0.61} & \multicolumn{2}{c}{$-$1.05} & \multicolumn{2}{c}{0.26} & \multicolumn{2}{c}{11.76} & \\
$z = 5$  & \multicolumn{2}{c}{0.98} & \multicolumn{2}{c}{$-$1.13} & \multicolumn{2}{c}{0.23} & \multicolumn{2}{c}{11.77} & \\
$z = 6$  & \multicolumn{2}{c}{0.77} & \multicolumn{2}{c}{$-$1.38} & \multicolumn{2}{c}{0.26} & \multicolumn{2}{c}{11.97} & \\
$z = 7$  & \multicolumn{2}{c}{0.82} & \multicolumn{2}{c}{$-$2.40} & \multicolumn{2}{c}{0.24} & \multicolumn{2}{c}{11.92} & \\
$z = 8$  & \multicolumn{2}{c}{0.64} & \multicolumn{2}{c}{$-$2.30} & \multicolumn{2}{c}{0.20} & \multicolumn{2}{c}{12.55} & \\
$z = 9$  & \multicolumn{2}{c}{0.55} & \multicolumn{2}{c}{$-$2.49} & \multicolumn{2}{c}{0.47} & \multicolumn{2}{c}{13.35} & \\
$z = 10$ & \multicolumn{2}{c}{0.78} & \multicolumn{2}{c}{$-$2.47} & \multicolumn{2}{c}{0.47} & \multicolumn{2}{c}{12.50} & \\
$z = 11$ & \multicolumn{2}{c}{0.95} & \multicolumn{2}{c}{$-$2.49} & \multicolumn{2}{c}{0.52} & \multicolumn{2}{c}{11.74} & \\
$z = 12$ & \multicolumn{2}{c}{1.23} & \multicolumn{2}{c}{$-$0.67} & \multicolumn{2}{c}{0.47} & \multicolumn{2}{c}{11.40} & \\
$z = 14$ & \multicolumn{2}{c}{2.76} & \multicolumn{2}{c}{$-$1.50} & \multicolumn{2}{c}{0.08} & \multicolumn{2}{c}{9.97} & \\
$z = 17$ & \multicolumn{2}{c}{2.23} & \multicolumn{2}{c}{$-$3.94} & \multicolumn{2}{c}{0.95} & \multicolumn{2}{c}{10.41} & \\
\toprule
\textbf{Conservative model} & \multicolumn{9}{c}{\textbf{Parameters}} \\
\textbf{With CS} & \multicolumn{9}{c}{} \\
\cmidrule(lr){2-10}
 & \multicolumn{2}{c}{$\alpha_\star$} & \multicolumn{2}{c}{$\beta_\star$} & \multicolumn{2}{c}{$\epsilon_{\star}$} & \multicolumn{2}{c}{$\log_{10}(M_c)$} & \hspace{1.5em}$G\mu$ \\
\midrule
$z = 4$  & \multicolumn{2}{c}{0.61} & \multicolumn{2}{c}{$-$1.04} & \multicolumn{2}{c}{0.20} & \multicolumn{2}{c}{11.94} & $7.59\times10^{-10}$  \\
$z = 5$  & \multicolumn{2}{c}{0.86} & \multicolumn{2}{c}{$-$0.90} & \multicolumn{2}{c}{0.23} & \multicolumn{2}{c}{11.82} & $1.05\times10^{-9}$ \\
$z = 6$  & \multicolumn{2}{c}{0.75} & \multicolumn{2}{c}{$-$1.41} & \multicolumn{2}{c}{0.23} & \multicolumn{2}{c}{12.03} & $1.07\times10^{-8}$ \\
$z = 7$  & \multicolumn{2}{c}{0.87} & \multicolumn{2}{c}{$-$2.53} & \multicolumn{2}{c}{0.21} & \multicolumn{2}{c}{11.81} & $8.13\times10^{-10}$ \\
$z = 8$  & \multicolumn{2}{c}{0.56} & \multicolumn{2}{c}{$-$0.82} & \multicolumn{2}{c}{0.20} & \multicolumn{2}{c}{12.64} & $3.63\times10^{-10}$ \\
$z = 9$  & \multicolumn{2}{c}{0.41} & \multicolumn{2}{c}{$-$1.65} & \multicolumn{2}{c}{0.61} & \multicolumn{2}{c}{14.10} & $1.26\times10^{-10}$ \\
$z = 10$ & \multicolumn{2}{c}{0.70} & \multicolumn{2}{c}{$-$2.45} & \multicolumn{2}{c}{0.44} & \multicolumn{2}{c}{12.57} & $5.13\times10^{-10}$ \\
$z = 11$ & \multicolumn{2}{c}{0.85} & \multicolumn{2}{c}{$-$2.40} & \multicolumn{2}{c}{0.47} & \multicolumn{2}{c}{11.92} & $7.76\times10^{-10}$ \\
$z = 12$ & \multicolumn{2}{c}{1.10} & \multicolumn{2}{c}{$-$2.46} & \multicolumn{2}{c}{0.50} & \multicolumn{2}{c}{11.62} & $9.54\times10^{-10}$ \\
$z = 14$ & \multicolumn{2}{c}{2.09} & \multicolumn{2}{c}{$-$2.89} & \multicolumn{2}{c}{0.06} & \multicolumn{2}{c}{12.26} & $3.50\times10^{-10}$ \\
$z = 17$ & \multicolumn{2}{c}{1.44} & \multicolumn{2}{c}{$-$3.90} & \multicolumn{2}{c}{0.43} & \multicolumn{2}{c}{10.90} & $9.58\times10^{-10}$ \\
\bottomrule
\end{tabular}}
\caption{\label{tab:paramsvalues} Values for the astrophysical parameters of the SFE and the cosmic string (CS) tension used for Figure \ref{fig:fitlowz} and Figure \ref{fig:fithighz}, as defined in Equation \ref{eq:conservativeparam} and in Equation \ref{eq:fiducialparam}}.
\end{table}

\newpage
\newpage

% Produces the bibliography via BibTeX.
% Produces the bibliography via BibTeX.
\end{document}